\documentclass{pasj01}
\Received{$\langle$reception date$\rangle$}
\Accepted{$\langle$acception date$\rangle$}
\Published{$\langle$publication date$\rangle$}
\usepackage{lscape}
\usepackage{subfigure}
\usepackage{multirow}
\usepackage{url}
\usepackage{threeparttable}
\usepackage{natbib}
\begin{document}

\title{Can the relativistic light bending model explain X-ray spectral variations of Seyfert galaxies?}
\author{Misaki Mizumoto$^{1,2}$, Kotaro Moriyama$^3$, Ken Ebisawa$^{1,2}$, Shin~Mineshige$^3$, Norita~Kawanaka$^{3,4}$, \& Masahiro~Tsujimoto$^1$}%
\altaffiltext{1}{Institute of Space and Astronautical Science (ISAS), Japan Aerospace Exploration Agency (JAXA), 3-1-1 Yoshinodai, Chuo-ku,
Sagamihara, Kanagawa, 252-5210, Japan}
\altaffiltext{2}{Department of Astronomy, Graduate School of Science, The University of Tokyo, 7-3-1 Hongo, Bunkyo-ku, Tokyo, 113-0033, Japan}
\altaffiltext{3}{Department of Astronomy, Kyoto University, Kitashirakawa-oiwake-cho, Sakyo-ku, Kyoto, 606-8502, Japan}
\altaffiltext{4}{Hakubi Center for Advanced Research, Kyoto University, Yoshida-hommachi, Sakyo-ku, Kyoto, 606-8501, Japan}
\email{mizumoto@astro.isas.jaxa.jp}


\maketitle

\begin{abstract}
Many Seyfert galaxies are known to exhibit Fe-K broad emission line features in their X-ray energy spectra.
The observed lines have three distinct features;
(1) the line profiles are skewed and show significant low-energy tails,
(2) the Fe-K band have low variability, which produces a broad and deep dip in the root-mean-square (rms) spectra,
and (3) photons in this band have time lags behind those in the adjacent energy bands with amplitudes of several $R_g/c$, 
where $R_g$ is the gravitational radius.
The ``relativistic light bending model'' is proposed to explain these observed features, 
where a compact X-ray source (lamp post) above an extreme Kerr black hole illuminates the innermost area of the accretion disc.
In this paper, we critically examine the relativistic light bending model by
computing the rms spectra and the lag features using a ray-tracing technique,
when a lamp post moves vertically on the black hole spin axis.
As a result, we found that the observed deep rms dip requires that the iron is extremely overabundant ($\gtrsim10$ solar),
whereas the observed lag amplitude is consistent with the normal iron abundance.
Furthermore, disappearance of the lag in the high-flux state requires a source height as high as $\sim40\,R_g$, which contradicts the relativistically broad emission line feature.
Our simulations agree with the data that the reverberation feature moves to lower frequencies with larger source height, however, if this scenario is correct, the simulations predict detection of a clear Fe-K lag at low frequencies, which is not constrained in the data.
Therefore, we conclude that the relativistic light bending model may not explain the characteristic Fe-K spectral variations in Seyfert galaxies.
\end{abstract}

%%%%%%%%%%%%%%%%%%%%%%%%%%%%%%%%%%%%%%%%%%%%%%%%%%%%%%%%%%%%%%%%%%%%%%%%%%%%
\section{Introduction} \label{sec1}

% Introduction of a broad line feature

Many Seyfert galaxies are known to show a ``broad iron emission line feature'' in their X-ray spectra around $\sim7$~keV (e.g., \citealt{tan95}).
When photons are scattered in neutral materials, such as outer part of the accretion disc or the dusty torus,
a sharp Fe-K fluorescent line is produced at 6.4~keV.
However, the observed iron line is commonly more broadened, often with a significant low-energy tail down to $\sim4$~keV.
If this feature is a real emission line,
it should be skewed by some physical mechanisms;
for example, \citet{fab89} and \citet{lao91} argued that the line broadening is caused by the gravitational redshift
due to the scattering on the disc very close to the central black hole (BH).
On the other hand, this spectral feature may be produced due to absorption features (e.g., \citealt{tan04,miz14,hag16}) or multi-continua \citep{nod11a}, 
depending on different continuum models.
These different mechanisms can equally explain the time-averaged spectral feature
at an accuracy of the contemporary instruments, and
thus origin of the broad iron line feature has still been under discussion.

Time-variability in the iron energy band should bring us key information to disentangle degenerate theoretical models.
Today, two major observational clues of time variability are reported: root-mean-square (rms) spectra and time-lags.
First, the rms spectra (energy dependence of the fractional variation) of Seyfert galaxies 
are known to have a deep and broad dip at $\sim4-7$~keV (e.g., \citealt{mat03,iso16,ter09}).
Namely, the Fe-K band in Seyfert galaxies have low variability compared to the continuum flux.
Depth of the dip is as large as $\sim50$\% relative to the reference energy band of $\sim2$~keV, red which is much larger than those expected solely due to the invariable and narrow neutral iron lines (e.g., \citealt{ter09}).
Second, photons in the Fe-K energy band are found to be delayed after those in the adjacent energy bands.
For example, \citet{zog12} and \citet{kar13b} found that the photons in 5--7~keV lag behind those in the adjacent energy bands in NGC 4151 and 1H 0707--495, respectively.
The lag amplitude corresponds to several $R_g/c$, where $R_g=GM_\mathrm{BH}/c^2$ is the gravitational radius and $M_\mathrm{BH}$ is the BH mass.
The broad feature is also seen in the lag-energy spectrum.
\citet{kar16} systematically investigated time lags of Seyfert galaxies in the {\it XMM-Newton} archive data, and
found that $\sim50$\% of sources have the Fe-K reverberation lags.
The lag amplitudes are commonly $1-9\,R_g/c$, and
the characteristic Fourier frequency is $\sim c/100\,R_g$~Hz.

The ``relativistic light bending model'' has been studied extensively to explain these observational facts.
In this model, relativistically-blurred reflection occurs at the innermost disc region around a Kerr BH with an almost maximum spin parameter (e.g., \citealt{fab02}).
When primary photons are scattered at the innermost region ($R\sim1.24\,R_g$),
strong gravitational redshift affects the reflected spectrum, and 
the Fe-K fluorescent line is skewed.
This model often assumes a compact corona ($\lesssim1\,R_g$), which is called as a ``lamp post'',
located just above the central BH with the source height of $h_s\lesssim10\,R_g$
illuminating the innermost region of the disc (e.g., \citealt{fab03,fab09}).
In order to explain the rms dips,
\citet{min03} and \citet{min04} considered a lamp post moving along the rotation axis.
When the source height is smaller, primary photons with a power-law spectrum are less likely to escape from the strong gravitational field in the vicinity of the BH while more likely to be trapped, so
the power-law component (PLC) becomes fainter.
On the other hand,
the reflection-dominated component (RDC) keeps almost invariant with a smaller $h_s$,
because larger parts of the bending photons fall into the BH itself and cannot reach the disc surface.
In this manner, variations of PLC and RDC are totally different, and
the variability amplitudes are expected to be reduced in the Fe-K energy band where RDC is dominant.
In addition,
the light-travel time from the lamp post to the accretion disc is about several $R_g$ (e.g., \citealt{kar13b,cac14}).
Therefore, the reverberation lags with short amplitudes ($\lesssim10\,R_g/c$) are expected.

Here, we point out potential problems in the relativistic light bending model.
One problem is that the rms spectra and lags seem to require different iron abundances.
The iron overabundance ($3-20$ solar) is commonly needed to explain the line profile and the rms spectra in the relativistic light bending model (e.g., \citealt{fab02,chi15}).
On the other hand, numerical simulations suggest that the solar iron abundance is sufficient to explain the observed lags \citep{emm14,cac14}.
Moreover, some argued that the observed deep rms dips cannot be fully explained only by the relativistic light bending model \citep{ino03,zyc10}.
If the relativistic light bending model is inconsistent, we may not trust physical parameters derived by this model.
For example, many papers publish a spin parameter, one of the three key parameters to characterise BHs, using this model via spectral fitting.
Therefore, it is very important to investigate validity of the relativistic light bending model.

In the previous studies, the two features about time variability in the Fe-K band, rms dips and reverberation lags, were explained individually, not simultaneously.
Therefore, in this paper, we perform a precise calculation of the relativistic light bending model
to study whether this model can explain the two features simultaneously.
The ray-tracing technique is adopted to compute photon paths under the strong gravitational field.
We select IRAS 13224--3809 as a representative object to investigate such Fe-K time features;
it has the strong Fe-K spectral feature, the deep rms dip, and the significant reverberation lags.
In addition, this source was observed with a very large program of {\it XMM-Newton}, and
we are benefitted by rich photon statistics to investigate detail of the spectral variations.
The iron line is strongly skewed (e.g., \citealt{bol03}),
the X-ray flux shows strong variability by more than one order of magnitude with a timescale of $\sim500$~s (e.g., \citealt{dew02}), and
the rms dip has an amplitude of $\sim50$\% (e.g., \citealt{yam16}). 
\citet{kar13c} also reported the Fe-K reverberation lags.
We try to explain all the Fe-K properties in this object.
First, we explain our models in \S\ref{sec2}, and
show the resultant rms spectra and lag features in \S\ref{sec3}.
Then we discuss whether the model can simultaneously explain the observed rms spectra and lag features of IRAS 13224--3809 in \S\ref{sec4}, and 
finally state our conclusions in \S\ref{sec5}.
%%%%%%%%%%%%%%%%%%%%%%%%%%%%%%%%%%%%%%%%%%%%%%%%%%%%%%%%%%%%%%%%%%%%%%%%%%%%
\section{Model and assumption} \label{sec2}

We briefly explain the model used in this paper
(see details in Appendix \ref{app1}).
We show in figure \ref{abs} the schematic picture for the model.
A distant observer observes the iron line reverberation due to a static X-ray emitter 
close to the prograde geometrically-thin accretion disc around a rotating BH.
The lamp post is assumed to emit isotropic X-ray pulses in its inertial frame.
The gas element of the accretion disc is postulated to be in Keplerian motion, and
the velocity components in the radial and polar directions are neglected. 
Radii of the inner and outer edges of the disc are fixed to be $r_{\rm ms}$ and $100\,R_{g}$, respectively, 
where $r_{\rm ms}$ is the radius of the marginally stable orbit.
The $r_{\rm ms}$ value is calculated as
\begin{eqnarray}
\frac{r_{\rm ms}}{R_g} &=& 3+Z_2 - \sqrt{(3-Z_1)(3+Z_1+2Z_2)}\\
& &Z_1 = 1+(1-a^2)^{1/3}\left\{ (1+a)^{1/3}+(1-a)^{1/3} \right\} \nonumber \\
& &Z_2 = \sqrt{3a^2+Z_1^2}, \nonumber
\end{eqnarray}
where $a=J/M_\mathrm{BH}^2$ is a spin parameter and $J$ is the angular momentum of the BH.
The light bending effects, Doppler effects, and gravitational redshift are taken into account.
The total mass of the disc is assumed to be negligible compared to the BH mass.

%%%%%%%%%%%%%%%%%%%%%%%%%%%%%%%%%%%%%%%%%%%
\begin{figure}
\begin{center}
\includegraphics[width=\columnwidth]{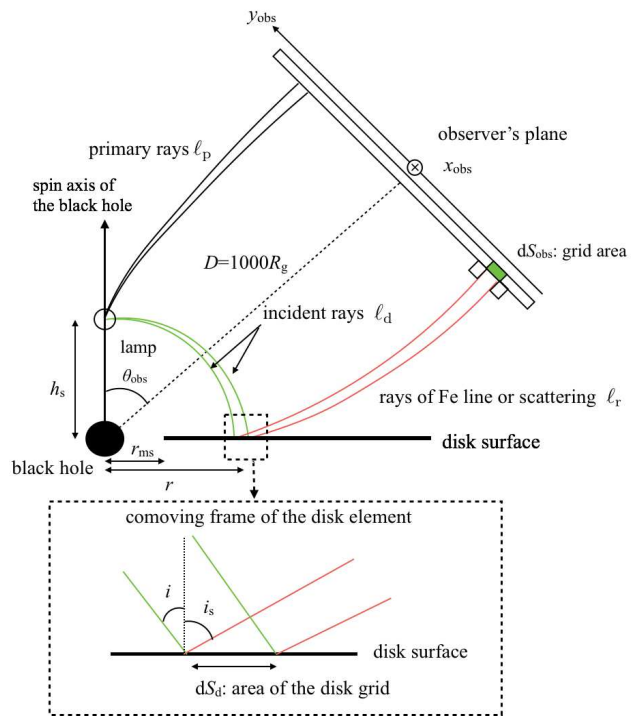} 
\end{center}
\caption{
Schematic picture of the relativistic light bending model. 
The X-ray emitter has the point-like geometry and radiates an isotropic X-ray flare in its inertial frame.
The accretion disc lies in the equatorial plane and is geometrically-thin and optically-thick.
The radius of the inner edge of the disc is equal to that of the marginally stable orbit, $r_{\rm ms}$.
Here, $h_{\rm s}$ is the height of the source, $\theta_{\rm obs}$ is the viewing angle, and $D$ is the distance between the BH and the centre of the observer's plane.
$\ell_{\rm p}, \ell_{\rm d},$ and $\ell_{\rm r}$ represent the photon path lengths of the primary, incident, and scattering orbits, respectively.
Area elements of the observer's plane and the disc surface are denoted as $dS_{\rm obs}$ and $dS_{\rm d}$, respectively.
In the comoving frame of the disc element, $i$ and $i_{\rm s}$ are the incident and scattering angles, respectively.
The $x_{\rm obs}$-axis is parallel to the equatorial plane of the BH ($\otimes$), and the $y_{\rm obs}$-axis is perpendicular to the $x_{\rm obs}$-axis.
}\label{abs}
\end{figure}
%%%%%%%%%%%%%%%%%%%%%%%%%%%%%%%%%%%%%%%%%%%

We numerically calculate trajectories of the photons emitted from
the static source with a point-like geometry.
Some photons directly reach the observer along the path $\ell_{\rm p}$,
which is regarded as the primary component.
Other photons illuminate the accretion disc along the path $\ell_{\rm d}$ 
and causes the fluorescent iron line and the reflected continuum emission,
which reach the observer along the path $\ell_{\rm r}$.
We assume the lamp post located on the rotation axis (e.g., model A in \citealt{nie10}).
For the ease of numerical calculation, intrinsic luminosity of the primary source is assumed to be invariable, and thus
only changes of $h_s$ above the disc induce observed variability.
Here, we point out that the rms spectral shape is invariable even if the intrinsic luminosity is variable,  
unless there is significant intrinsic {\it spectral} variations.
The observed spectra were computed for different $h_s$.
The heights ($h=h_s/R_g$) were linearly spaced with $\Delta h=1$ at $h\geq3$ and
$\Delta h=0.2$ at $h<3$, following \citet{nie10}.
The $h$ range was set to be $2.2 \leq h \leq 10$.
Calculations are performed for four types of spin parameters ($a=0,\,0.6,\,0.9,\,0.998$).
The photon index of the primary power-law spectrum is fixed at $2$.
We assume a high inclination angle ($i=60$~deg), which is applicable for IRAS 13224--3809 \citep{bol97,par17a}.
Calculations for the low-inclination case ($i=30$~deg) are described in Appendix \ref{app2}.
%%%%%%%%%%%%%%%%%%%%%%%%%%%%%%%%%%%%%%%%%%%%%%%%%%%%%%%%%%%%%%%%%%%%%%%%%%%%
\section{Results} \label{sec3}

\subsection{Energy spectra}

\begin{figure*}
\centering
\subfigure{
\resizebox{8cm}{!}{\includegraphics[width=60mm,angle=270]{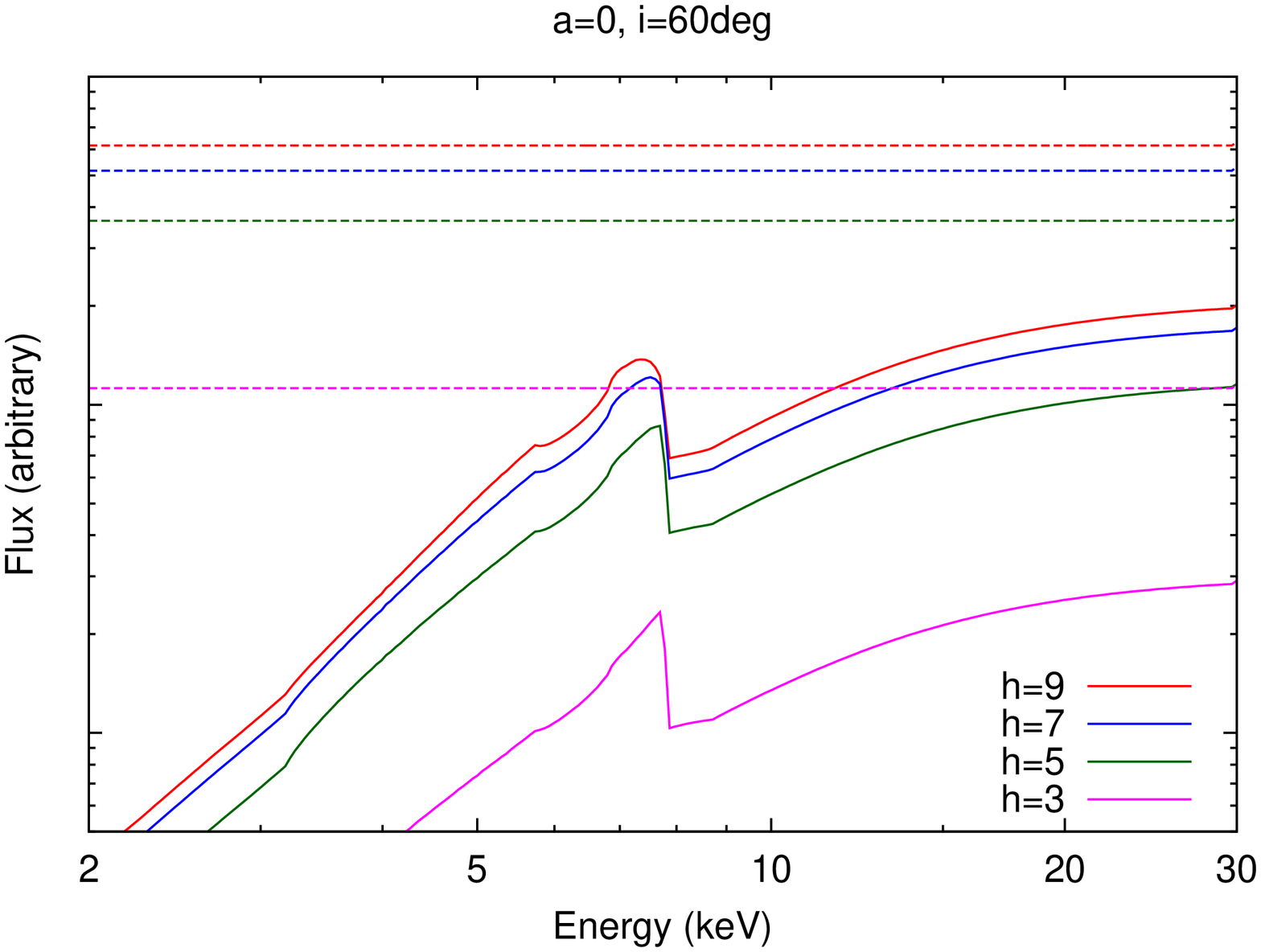}}
\resizebox{8cm}{!}{\includegraphics[width=60mm,angle=270]{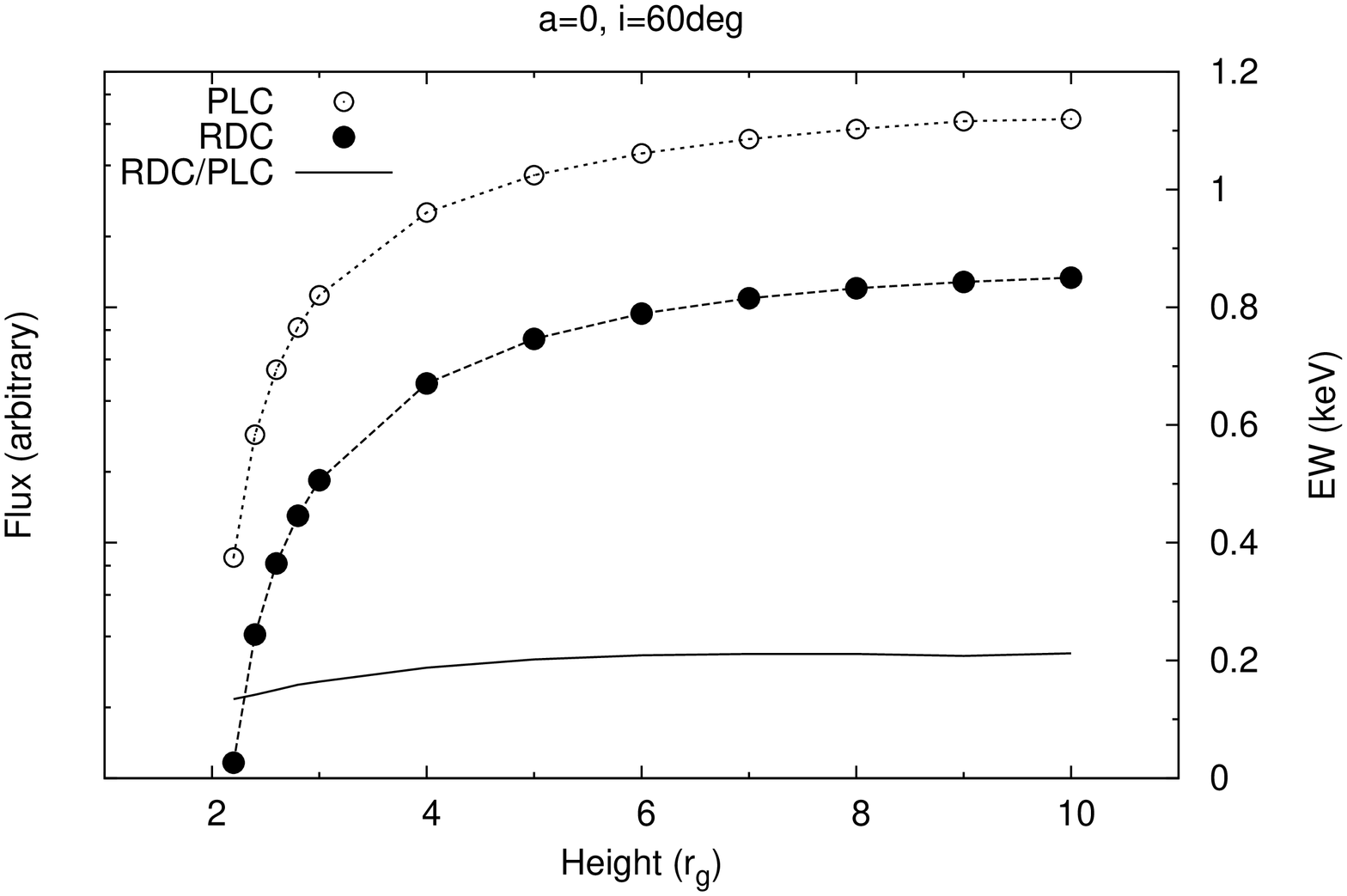}}
}
\subfigure{
\resizebox{8cm}{!}{\includegraphics[width=60mm,angle=270]{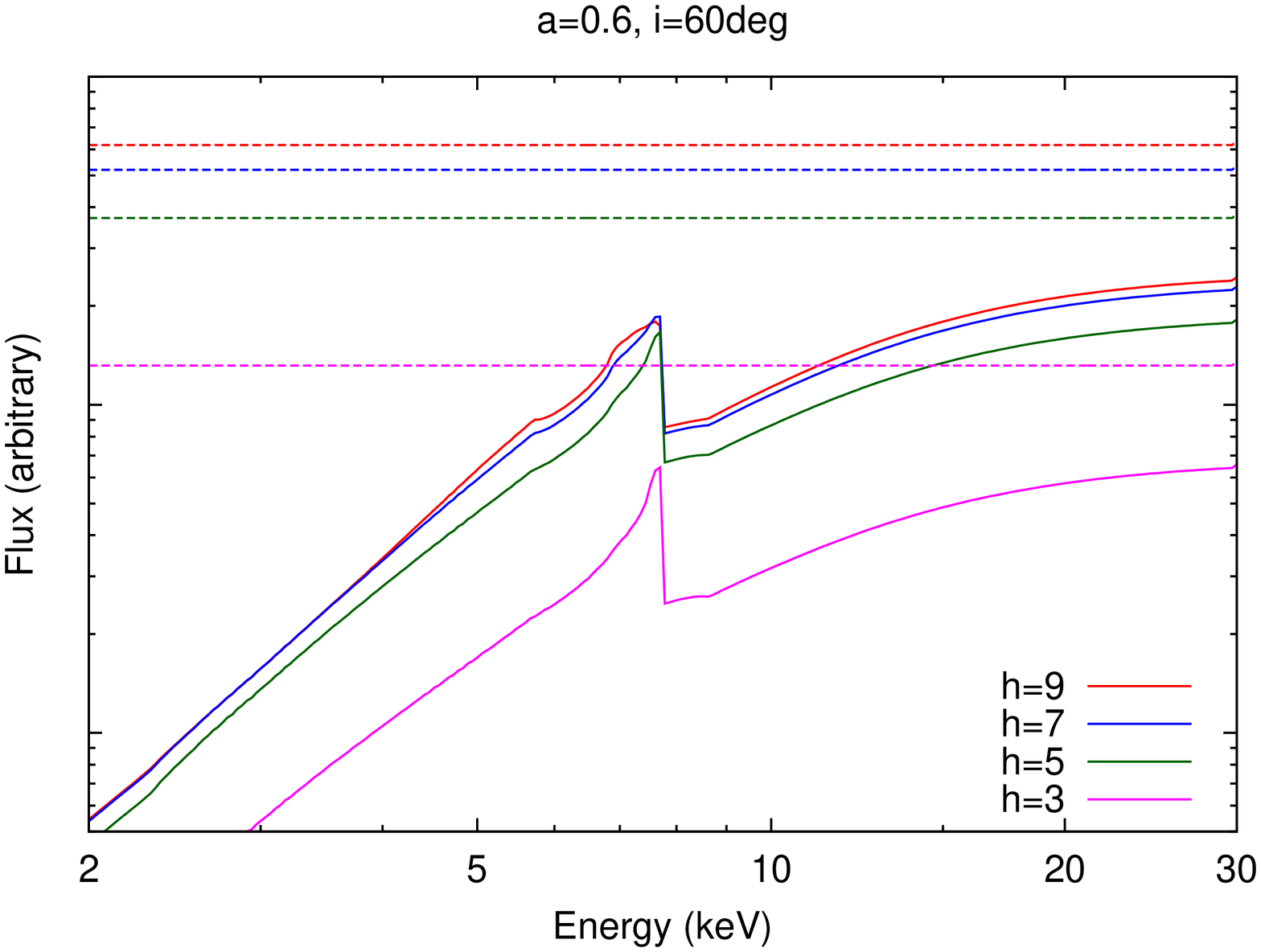}}
\resizebox{8cm}{!}{\includegraphics[width=60mm,angle=270]{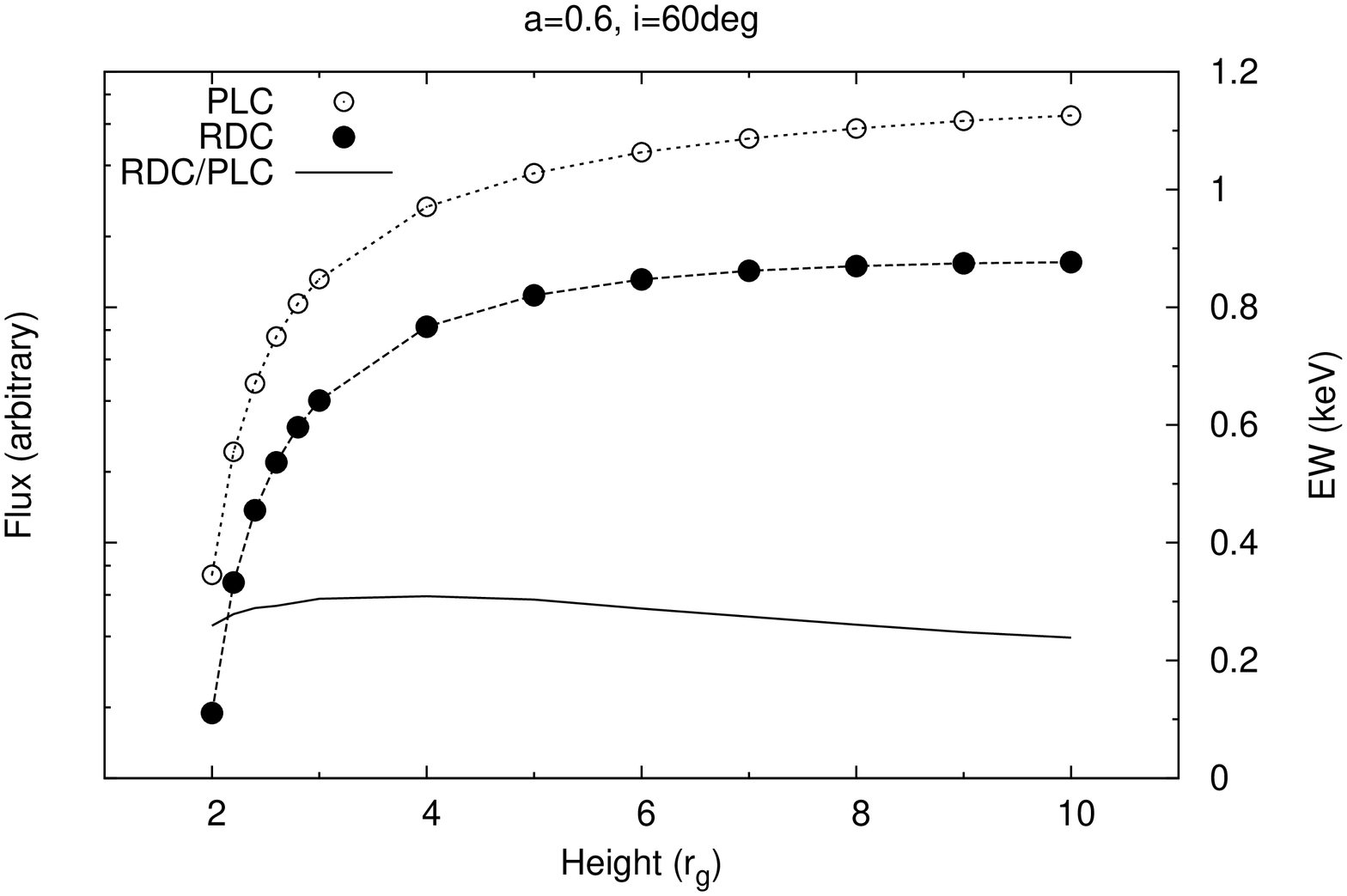}}
}
\subfigure{
\resizebox{8cm}{!}{\includegraphics[width=60mm,angle=270]{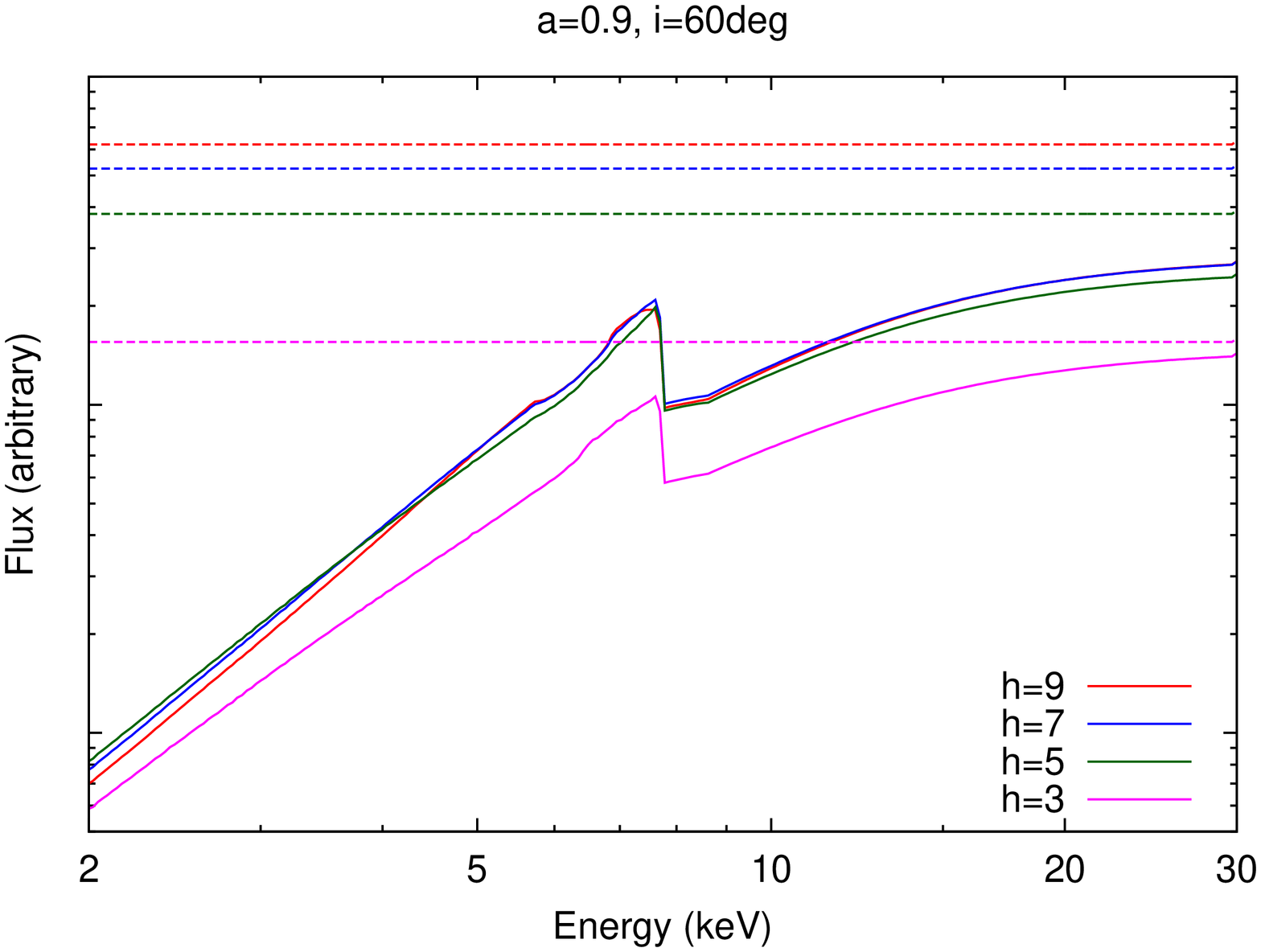}}
\resizebox{8cm}{!}{\includegraphics[width=60mm,angle=270]{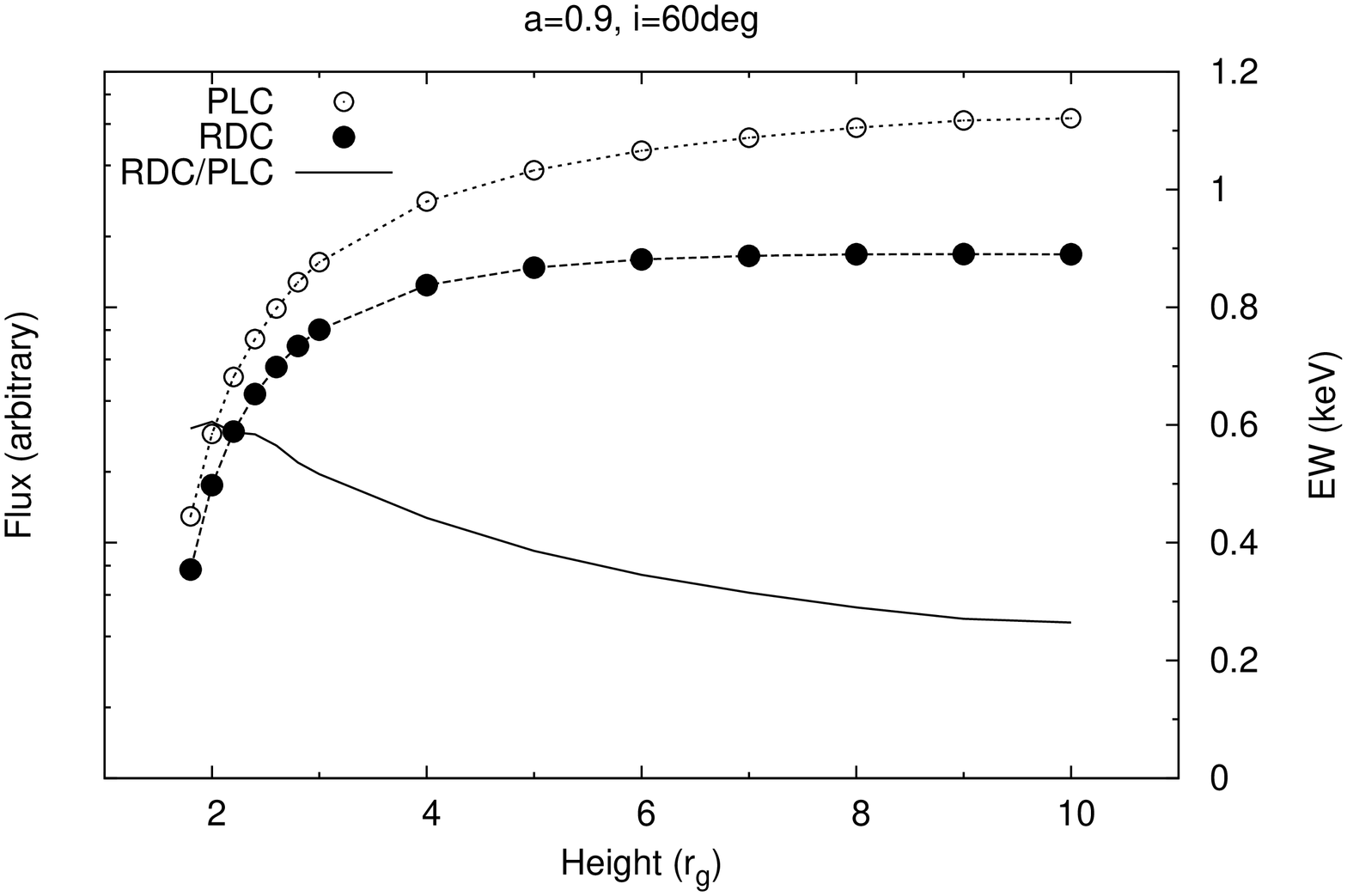}}
}
\subfigure{
\resizebox{8cm}{!}{\includegraphics[width=60mm,angle=270]{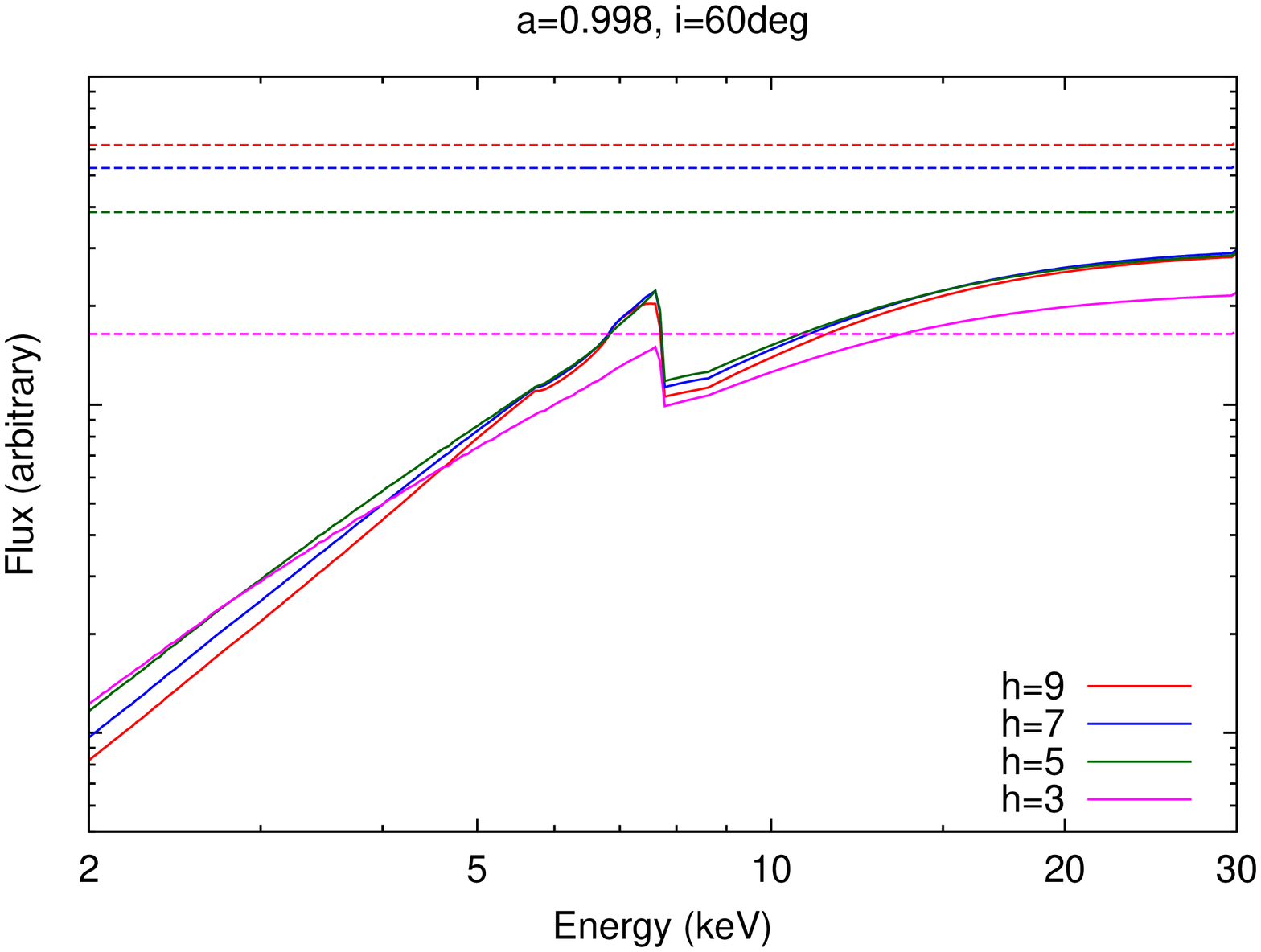}}
\resizebox{8cm}{!}{\includegraphics[width=60mm,angle=270]{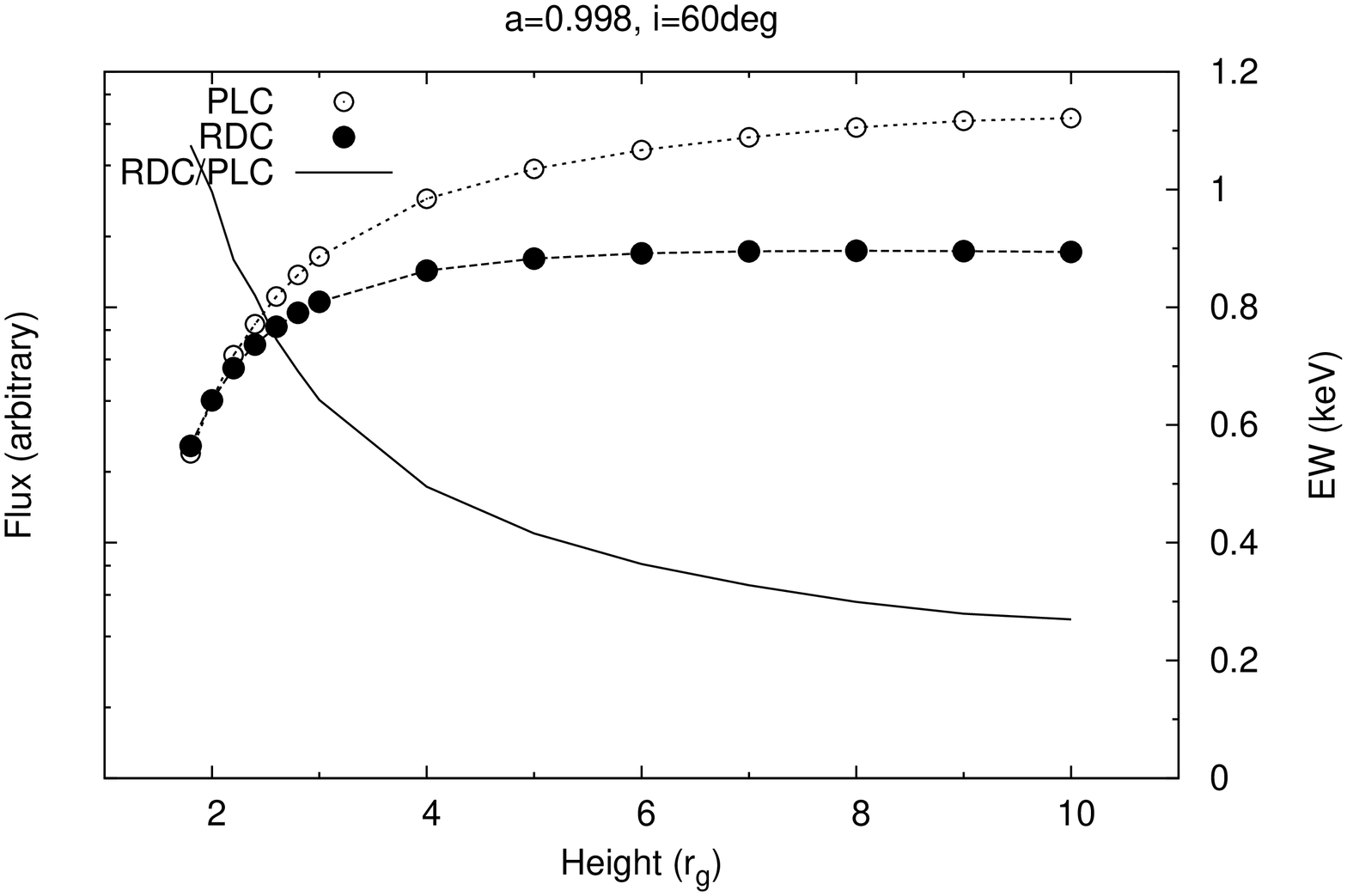}}
}
\caption{
(Left) Primary and reflected (iron line + Compton reflection) spectra for different $a$ and $h$, shown by dashed and solid lines, respectively.
The vertical axis shows the energy flux (an arbitrary unit).
(Right) The PLC (flux at 6.4~keV) and RDC (integrated over the iron line) fluxes as a function of $h$.
The iron line EW $\equiv$ RDC/PLC is also shown in the solid line. 
Units of flux are arbitrary, whereas that of EW is given in keV.
}
\label{fig:ironspectra}
\end{figure*}

Figure \ref{fig:ironspectra} shows $h$ dependence of the primary and reflected spectra for different $a$.
Find the red lines in the top left panel for explanation ($a=0$ and $h=9$ case):
The horizontal line shows the primary spectrum, whereas the curved one shows the reflected spectrum.
In the reflected spectrum, Fe-K lines, as well as the absorption edges, are skewed due to the gravitational redshift and the Doppler effect.
As $h$ decreases (see other colours in the same panel), 
the effect of the gravitational redshift becomes stronger, fewer photons escape from the gravitational field,
and both the observed iron line fluxes and the primary component become smaller.
On the other hand, Doppler effect of the disc has little dependence on $h$, 
such that the cut-off energies of the blue lines are unchanged. 
When $a$ increases (see red lines in the other panels), the inner edge of the disc moves inward and 
the line is more strongly skewed by the gravitational redshift.
Therefore, the low-energy tail reaches the softer band and the Fe-K emission line becomes broader.

The right panel of figure \ref{fig:ironspectra} shows variations of the PLC flux, the RDC flux, and the equivalent widths (EW) of the iron line ($=$RDC flux/PLC flux),
where the PLC flux shows the 6.4~keV flux and the RDC flux shows the integrated flux over the iron line \citep{min04}.
When $a\lesssim0.6$, variations of the RDC and PLC fluxes are similar, so
the EW is rather flat against significant change of $h$.
In the larger spin cases ($a>0.9$), 
the RDC flux is almost flat due to the light-bending effect in the $h$ range of $\sim5-10$ while the PLC flux is more significantly variable;
this is considered to be the cause of the Fe-K dip in the rms spectrum.

\subsection{Spectral variability}
We compute 12 simulated spectra for different $h$ ($=2.2-10$) and calculated fractional variability amplitudes ($F_\mathrm{var}$; \citealt{ede02}) using the following equation:
\begin{equation}
F_\mathrm{var}(E)=\frac{1}{\langle X \rangle}\sqrt{S^2-\langle\sigma^2_\mathrm{err} \rangle},
\end{equation}
where $X_i$ is the photon count in the energy bin of interest for the $i$-th among the 12 simulated spectra, 
$\langle X \rangle$ is the mean counts, 
$S^2$ is the variance of $\{X_i\}$, and 
$\langle\sigma^2_\mathrm{err} \rangle$ is the mean error squared of $\{X_i\}$, which is assumed to be null in our calculations.
We assume that $h$ varies continuously between 2.2 and 10.
We also calculate the rms spectra assuming different variation pattern of $h$ (such that $h$ variation follows the sinusoidal function), only to find that the resultant rms features hardly change.

\begin{figure*}
\begin{center}
\subfigure{
\resizebox{8cm}{!}{\includegraphics[width=60mm,angle=270]{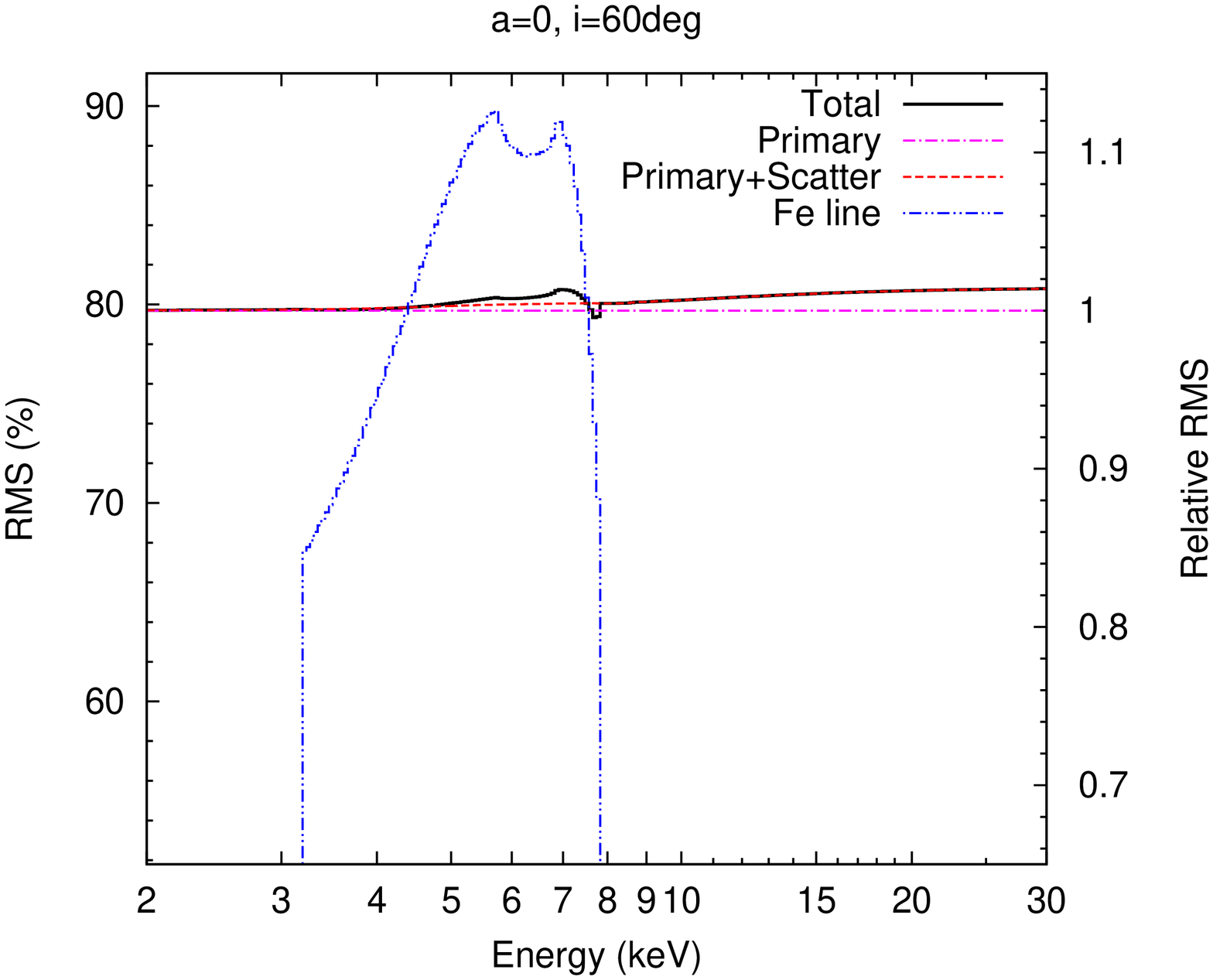}}
\resizebox{8cm}{!}{\includegraphics[width=60mm,angle=270]{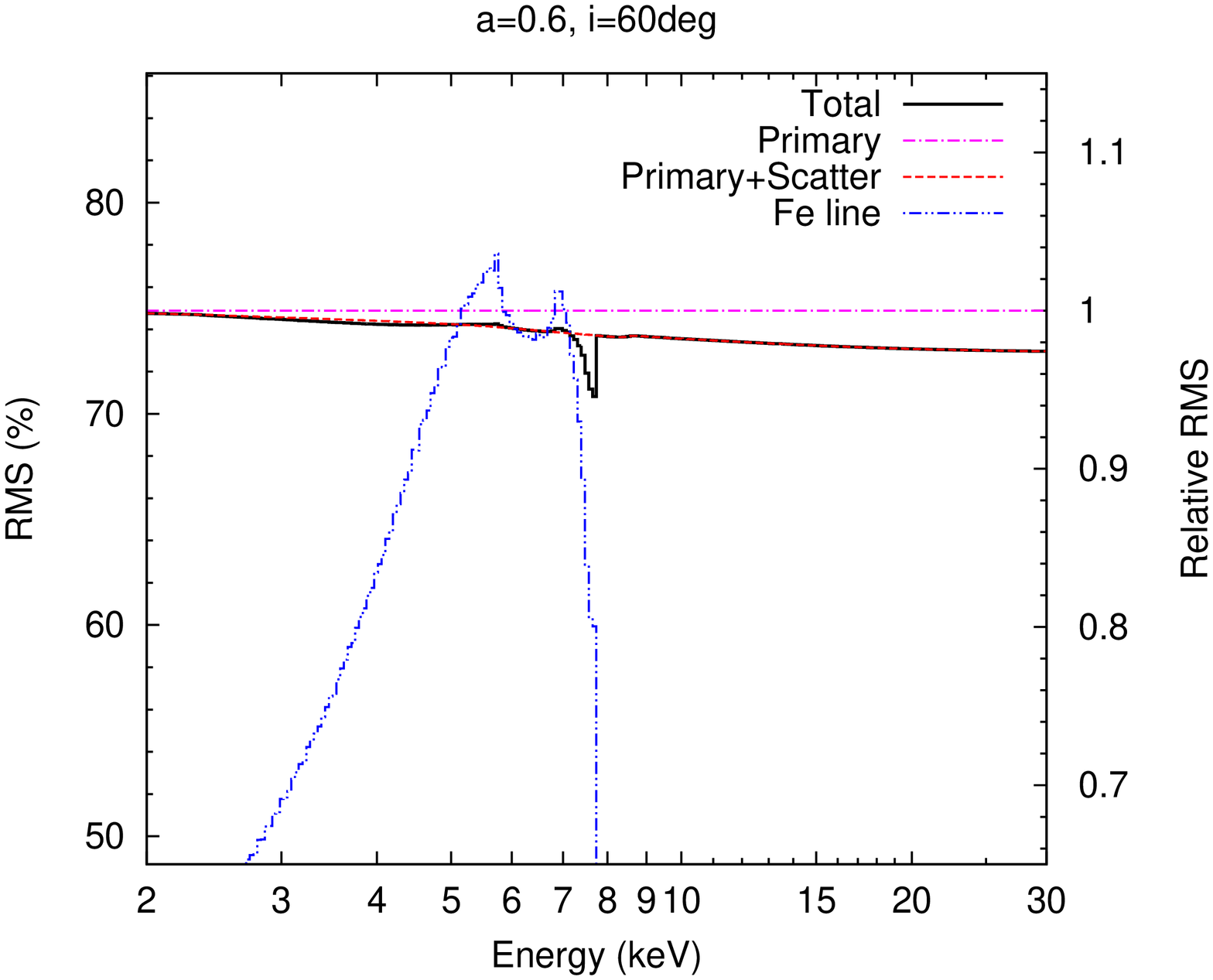}}
}
\subfigure{
\resizebox{8cm}{!}{\includegraphics[width=60mm,angle=270]{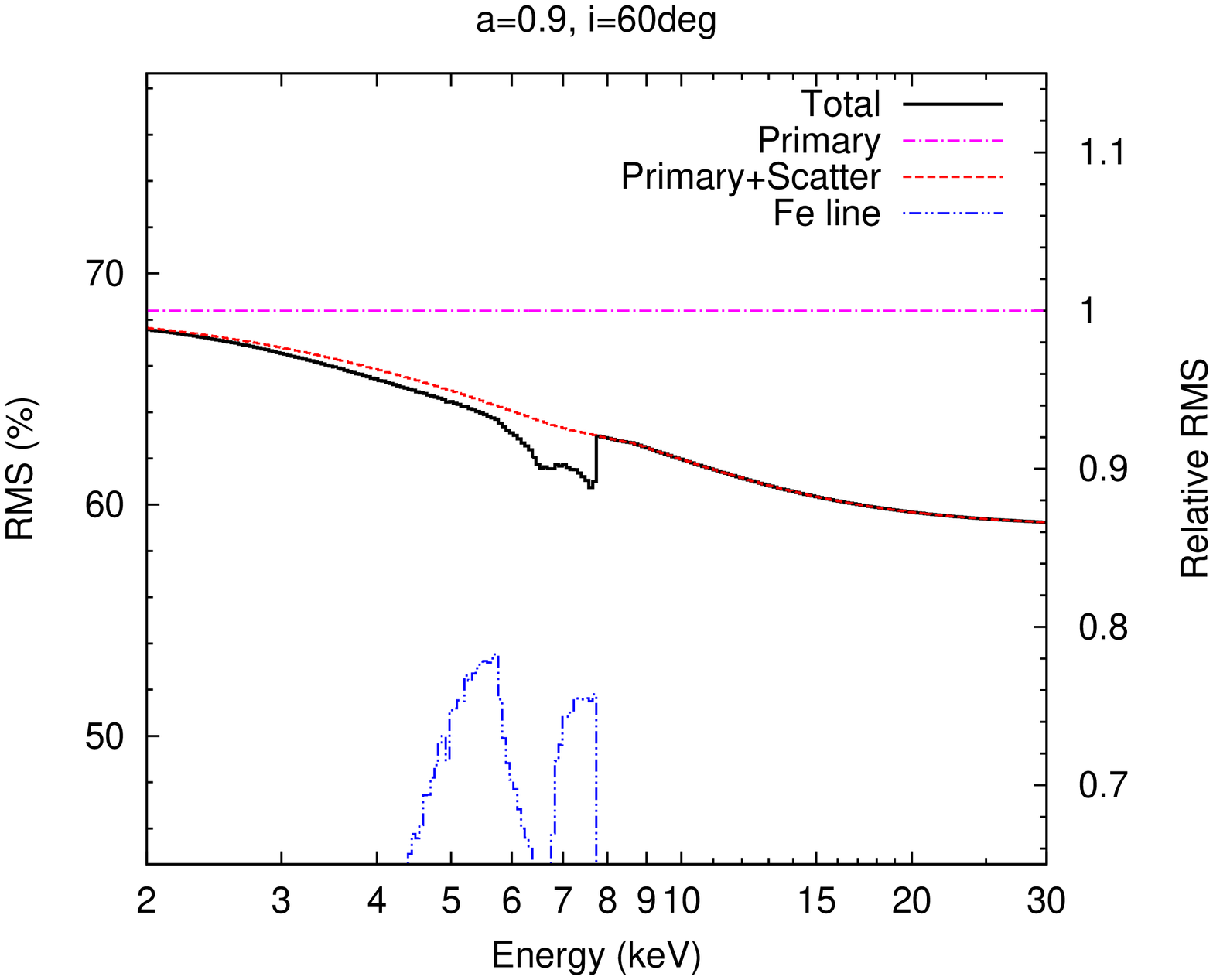}}
\resizebox{8cm}{!}{\includegraphics[width=60mm,angle=270]{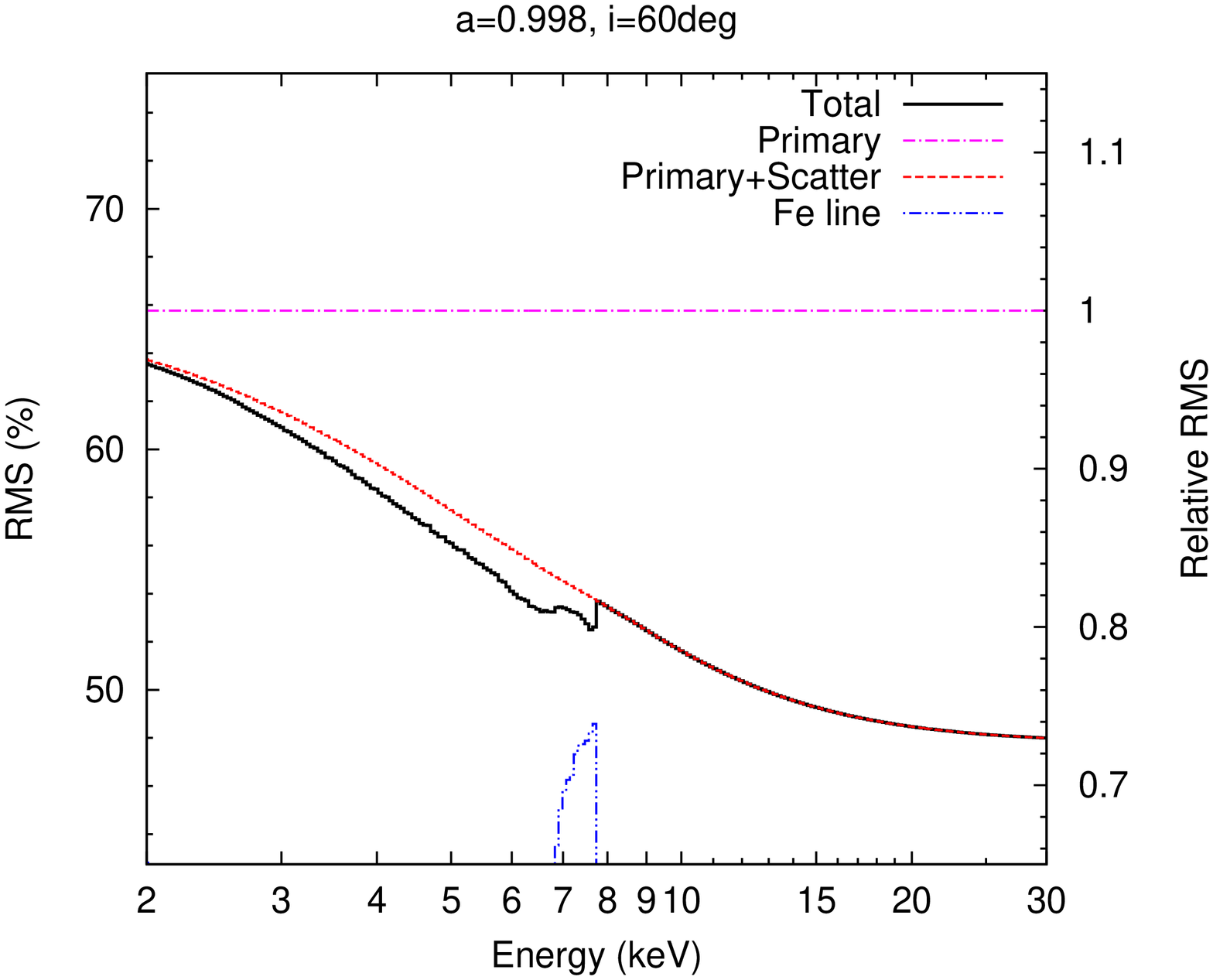}}
}
\caption{
The rms spectra for different spectral components.
The curves are for the primary component (magenta dot-dashed),
the primary$+$Compton reflection (red dashed), and the iron line (blue dot-dot-dashed).
The black solid curve indicates the total emission (primary$+$Compton reflection$+$iron line).
The right axis show the relative amplitude where that of primary component is normalised as unity.
}
\label{fig:rms}
\end{center}
\end{figure*}

Figure \ref{fig:rms} shows the resultant rms spectra.
In the zero spin case, 
variations of PLC and RDC are similar and EW of the iron line is constant against different $h$ (figure \ref{fig:ironspectra}),
so the rms spectrum is rather featureless.
In the spinning case, RDC varies less than PLC  (figure \ref{fig:ironspectra}), and
both Compton reflection and iron line reduce variation amplitudes.
In particular, the rms dip gets deeper and broader for the higher spin ($a\geq0.9$).
The relative dip depths at the 7--8~keV band are up to $20$\% in the maximum spin case.

\subsection{Time lags}\label{sec:lag}

\begin{figure*}
\begin{center}
\includegraphics[width=150mm]{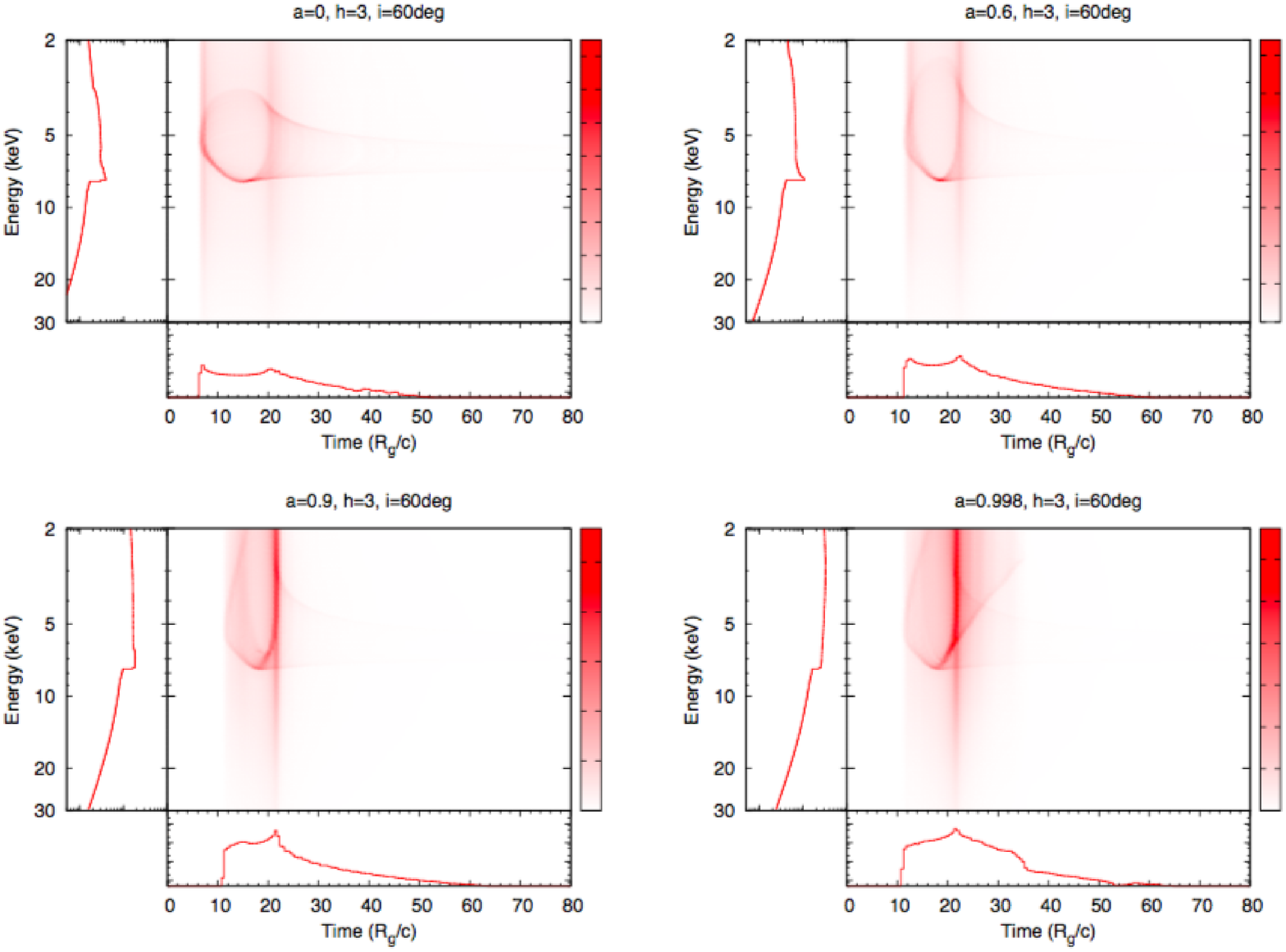}
\caption{
Main panels: Counts of the photons reflected on the disc 
for different $i$ and $a$ when $h = 3$.
The X-ray source emits at once with a delta-function of time.
The color bar shows the photon counts in the arbitrary unit.
Left-hand panels: Projection to the $y$-axis, which gives the total energy spectrum of all the reflected photons. 
Bottom panels: Projection to the $x$-axis, which gives the energy-averaged response
}
\label{fig:2dtf}
\end{center}
\end{figure*}

Next, we compute time lags (also see \citealt{miz18}).
Here, the time lag is defined as
\begin{equation}
\tau(f)=\mathrm{arg}[\mathcal{S}(f)\mathcal{H}^*(f)]/(2\pi f) \label{eq1},
\end{equation}
where $^*$ means complex conjugate and $\mathcal{S}(f)$ and $\mathcal{H}(f)$ are 
Fourier transforms of soft- and hard-band light curves, $s(t)$ and $h(t)$ \citep{vau97,now99}.  
We use the 3--4 keV (less contaminated by the reflected component) for the soft band
and 5--8 keV (dominated by the reflected one) for the hard band.
A positive lag means the hard band lagging the soft, and vice versa.
An observed light curve is expressed as 
the sum of the primary emission and the reprocessed emission.
Now we assume that the primary emission is expressed as $P(E)g(t)$,
where $P(E)$ is the spectrum of the primary component and $g(t)$ is the intrinsic flux variability.
We define $R(E,k)$ as photon counts of the reflected component in each energy-bin ($E$) and time-bin ($k$).
In this setting, the light curves are written as
\begin{eqnarray}
s(t)&=&P_s g(t) + \sum_{k} R_s(k) g(t- k t_{\rm bin}) \\
h(t)&=&P_h g(t) + \sum_{k} R_h(k) g(t- k t_{\rm bin}),
\end{eqnarray}
where $t_\mathrm{bin}$ is a time bin-size of the light curve and $P_{s,h}$ and $R_{s,h}(E)$ are the primary components and the reflected components, respectively, and
the suffixes indicate the energy bands.
Their Fourier transforms are expressed as
\begin{eqnarray}
\mathcal{S}(f)&=&P_s \mathcal{G}(f) + \sum_k R_s(k)\exp[-2\pi i(kt_\mathrm{bin})f]\mathcal{G}(f) \\
\mathcal{H}(f)&=&P_h \mathcal{G}(f) + \sum_k R_h(k)\exp[-2\pi i(kt_\mathrm{bin})f]\mathcal{G}(f), \label{eq:Fourier}
\end{eqnarray}
where $\mathcal{G}(f)$ are Fourier transform of $g(t)$.
Unless $\mathcal{G}(f)=0$, equation (\ref{eq1}) is calculated as
\begin{eqnarray}
\tau(f)&=&\frac{1}{2\pi f}\;\mathrm{arg}\left[\left(P_s+\sum_k R_s(k)\exp[-2\pi i(kt_\mathrm{bin})f]\right)\right.\nonumber\\
&& \:\:\:\:\:\:\:\:\:\:\:\:\:\:\:\:\:\:\:\:\:\:\:
\left.\left(P_h+\sum_k R_h(k)\exp[2\pi i(kt_\mathrm{bin})f]\right)\right]. \label{eq:lagdef}
\end{eqnarray}
Note that {\it this equation does not depend on the
functional form of the intrinsic variation}.
$P(E)$ and $R(E,k)$ are directly calculated from our simulations.
$P(E)$ corresponds to the primary component in figure \ref{fig:ironspectra}.
Figure \ref{fig:2dtf} shows $R(E,k)$, i.e., counts of the reflected photons in each time- and energy-bin.
This corresponds to the two-dimensional (2D) transfer function applied to the input photon spectrum.
Interpretation of the 2D transfer functions are described in previous papers (e.g.,\,\citealt{rey99, wil13, cac14}).
The delay time of a scattered photon is determined by the photon path difference from the primary photons, as well as Shapiro delays.
The main difference among different $a$ is that 
the energy-averaged responses in the high-spin cases have stronger peaks than in the low-spin cases.
This difference is because the disc with high $a$ has a larger surface area than that with low $a$ and extends closer to the BH,
whereas the response from the outer part of the disc is still the same (also see \citealt{cac14}).

\begin{figure*}
\begin{center}
\subfigure{
\resizebox{8cm}{!}{\includegraphics[width=65mm,angle=270]{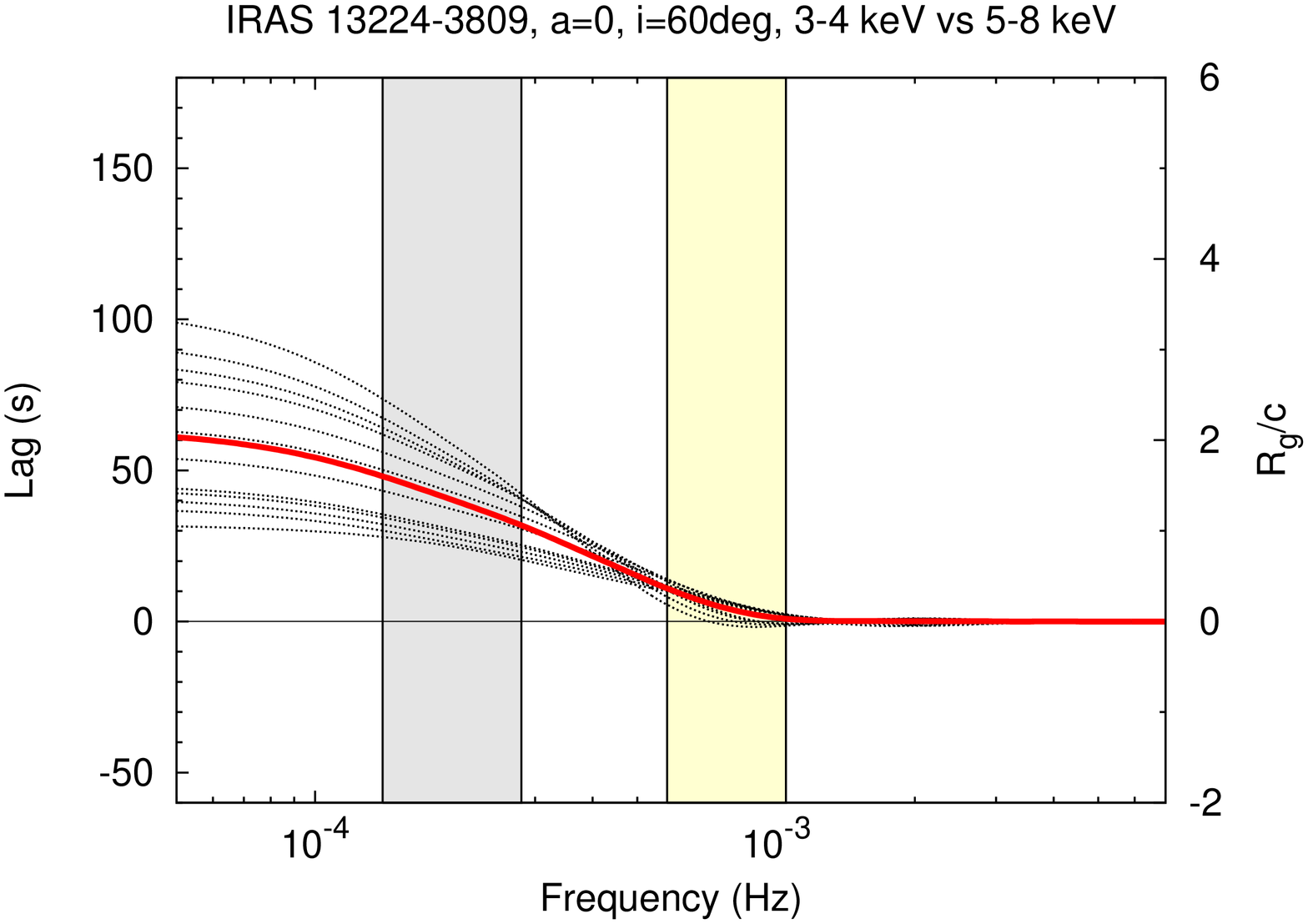}}
\resizebox{8cm}{!}{\includegraphics[width=65mm,angle=270]{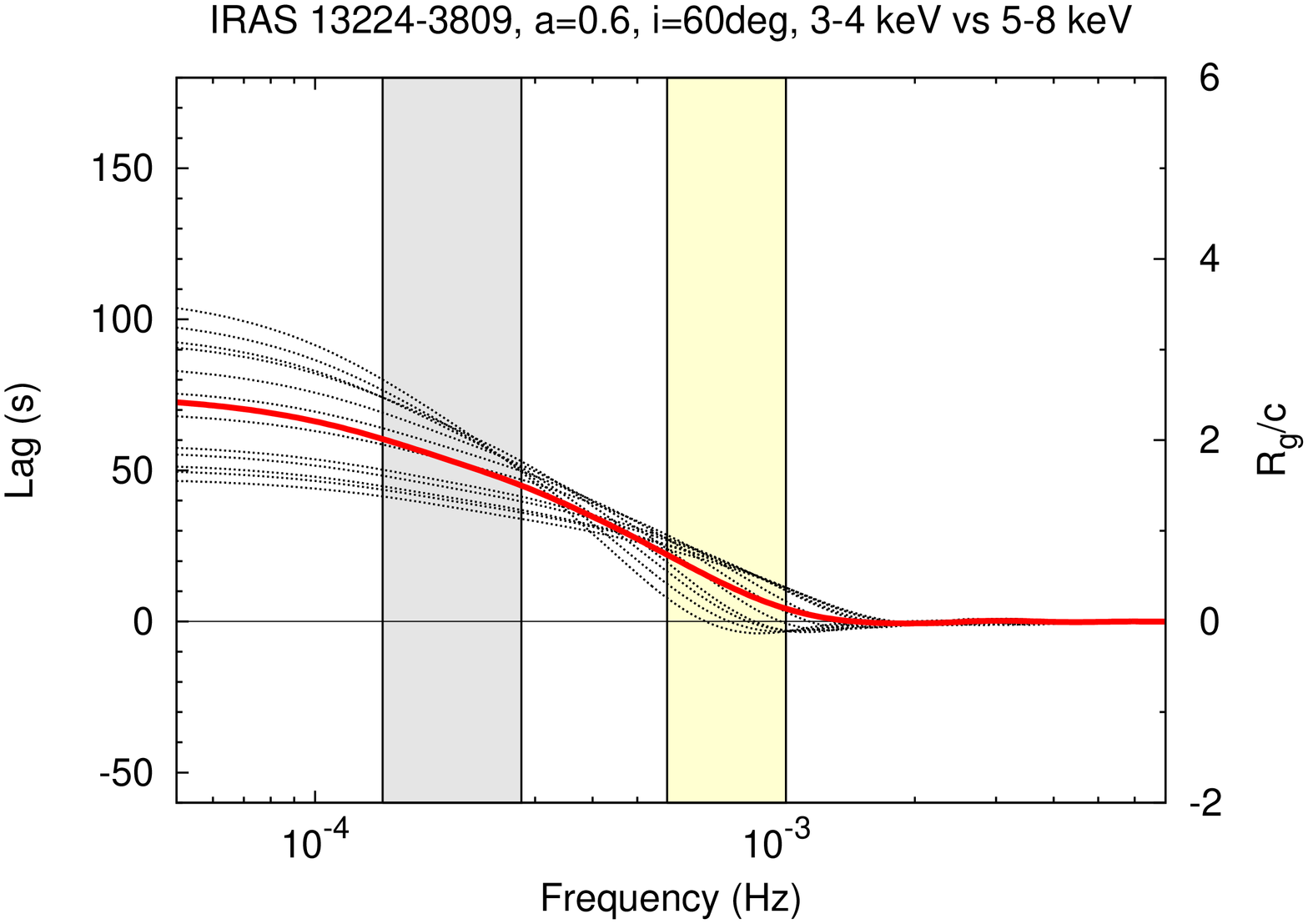}}
}
\subfigure{
\resizebox{8cm}{!}{\includegraphics[width=65mm,angle=270]{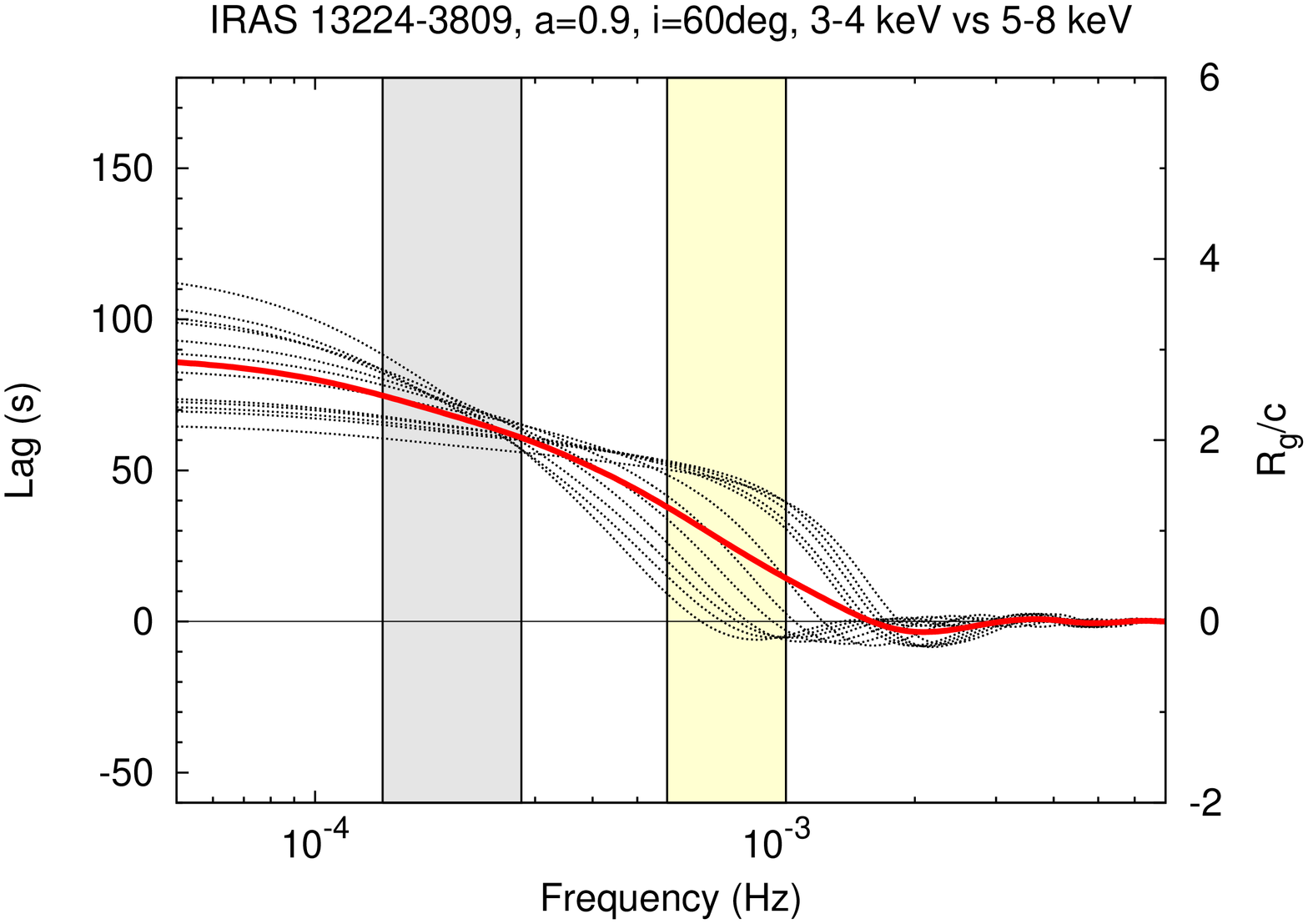}}
\resizebox{8cm}{!}{\includegraphics[width=65mm,angle=270]{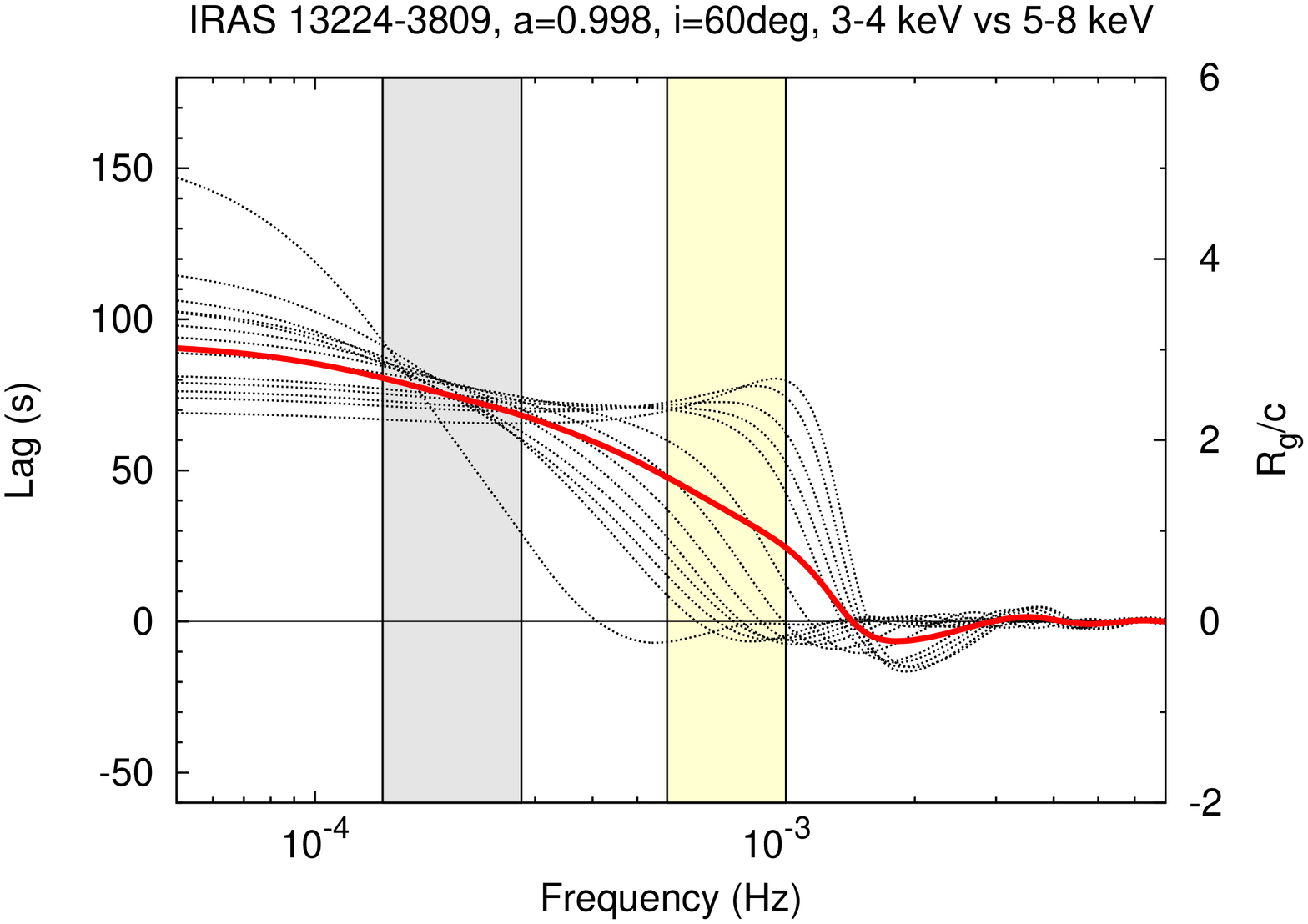}}
}\caption{
Lag-frequency plots, comparing 3--4 keV with 5--8 keV, with $h$ values of 2.2, 2.4, 2.6, 2.8, 3.0, 4.0, 5.0, \dots, 10.0 from bottom to top at the lowest frequency.
A positive lag means that the hard band lags behind the soft.
The BH mass is $10^{6.8}\,M_\odot$, i.e., $R_g/c=30$~s, assuming IRAS 13224--3809.
The yellow area shows $[5.8-10.5]\times10^{-4}$~Hz where the Fe-K lag is prominently seen in IRAS 13224--3809, 
and the grey shows $[1.4-2.8]\times10^{-4}$~Hz where the lag disappear in the high-flux periods of the same source.
The red solid line shows the average plot in $h=2.2-10$.
The plot of $h=20$ is additionally shown only in the $a=0.998$ case.
}
\label{fig:lagf}
\end{center}
\end{figure*}

First, we calculate the lags for each fixed $h$ value.
We assume that the BH mass is $10^{6.8}\,M_\odot$ \citep{gon12}, i.e., $R_g/c=30$~s.
The black dotted lines in figure \ref{fig:lagf} shows the lag-energy plots of each $h$ value.
The lag amplitude at the lowest frequency decreases with lower $h$ corresponding to the light-travel time, but
above the pivot point at $\sim[1-5]\times10^{-4}$~Hz, the lag amplitude increases.
Next, with the probability distribution of $h$ being flat within the variation range of $h=2.2-10$,
we calculate the average lag-frequency plot (red-solid line in figure \ref{fig:lagf}).
The lag amplitude in the high frequency range ($\gtrsim5\times10^{-4}$~Hz) is larger for larger $a$ 
because reflection on the innermost part of the disc is stronger.

\begin{figure}
\centering
\includegraphics[width=70mm,angle=270]{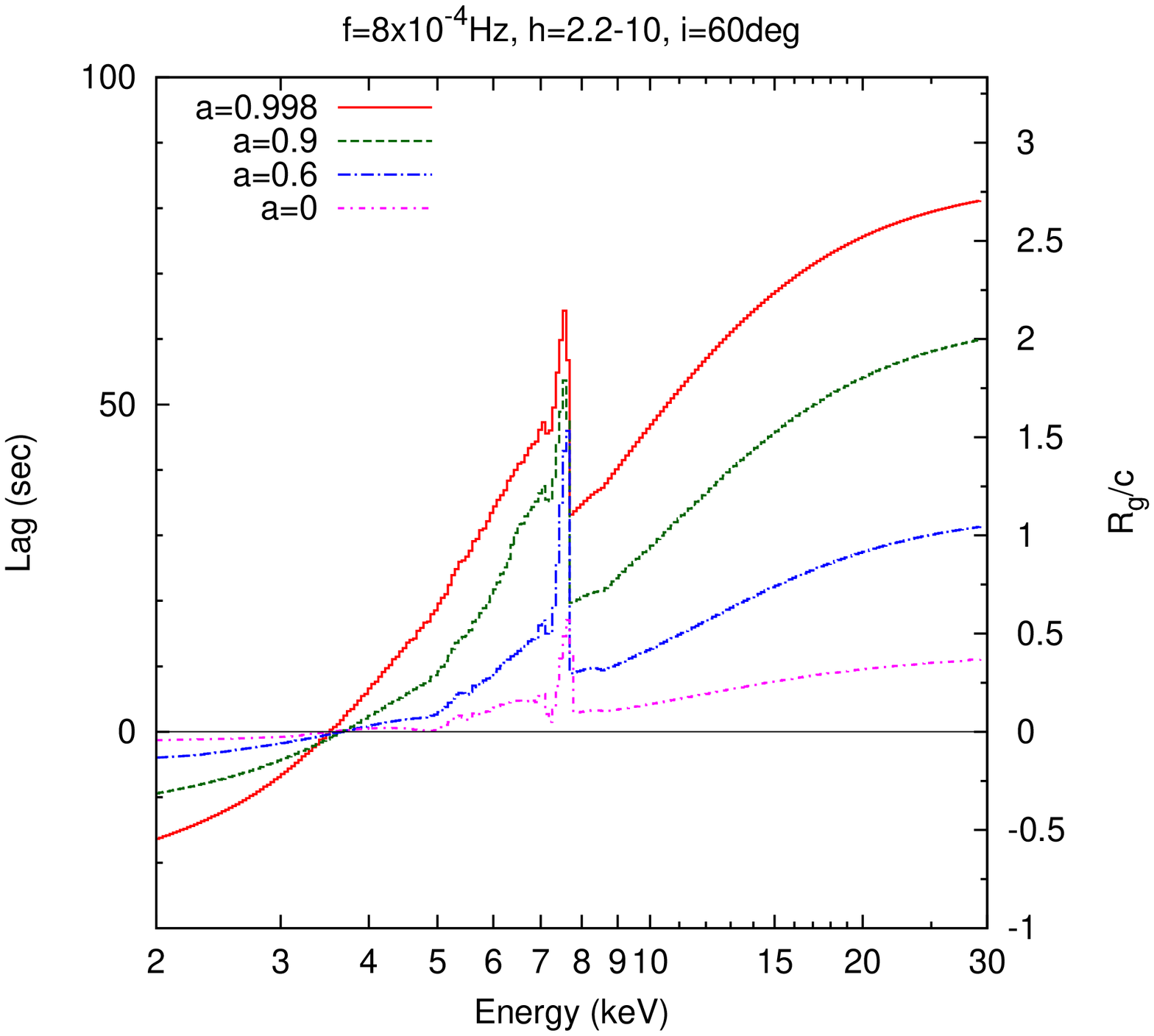}
\caption{
Lag-energy plots for different $a$ at the frequency of $8\times10^{-4}$~Hz
}
\label{fig:lagE}
\end{figure}

Figure \ref{fig:lagE} shows the lag-energy spectra for different $a$ in the high frequency range ($8\times10^{-4}$~Hz).
We compute a lag between each energy band of interest and a broad reference band of 2--30~keV.
All energy bins include both the primary and disc-reflection components, regardless of which dominates in the bins.
This means that the dilution effects are fully taken into account.
The reverberation Fe-K lags are seen in all the cases, but
the broad features in the 4--8~keV band are only seen when $a>0.9$.
The lag amplitude is in the range of 1--2~$R_g/c$.

%%%%%%%%%%%%%%%%%%%%%%%%%%%%%%%%%%%%%%%%%%%%%%%%%%%%%%%%%%%%%%%%%%%%%%%%%%%%
\section{Discussion} \label{sec4}

\subsection{Comparison with observations of IRAS 13224--3809} 
Our calculations show that the Fe-K rms dip ($\lesssim20$\%) and
the broad Fe-K lag with the amplitude of $\sim50$~s are seen when $a>0.9$.
Here, we carefully investigate whether these results can quantitatively explain observations of IRAS 13224--3809.

\subsubsection{rms spectra}
\begin{figure}
\centering
\includegraphics[width=60mm,angle=270]{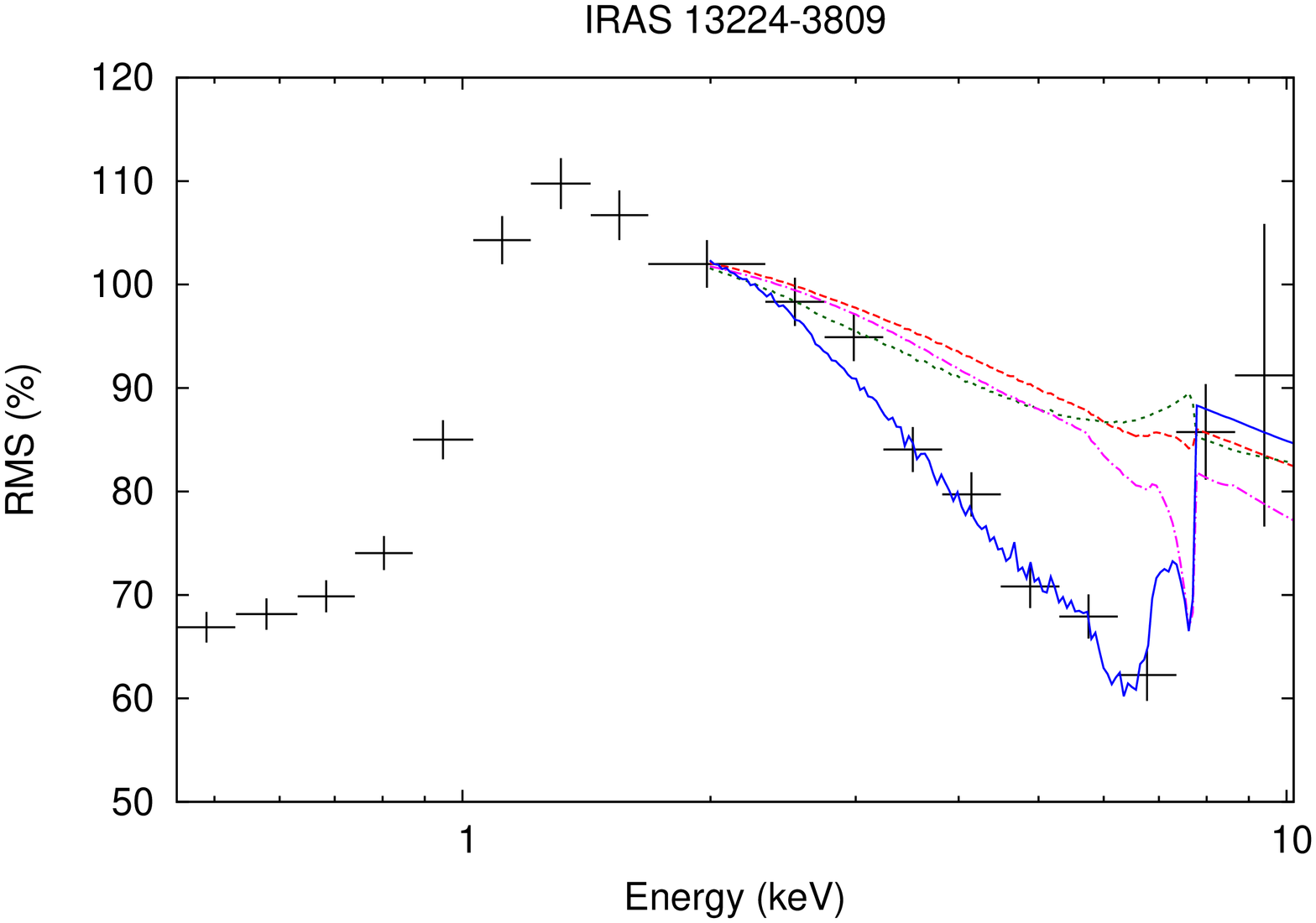}
\caption{
Observed rms spectra of IRAS 13224--3809 (black-solid bins) compared with the relativistic disc reflection model with $a=0.998$, normalised at 2 keV.
The red dashed, green dotted, and magenta dot-dashed lines are for the solar abundance with $2\leq h\leq10$, $2\leq h\leq3$, and $5\leq h\leq10$, respectively.
The blue solid line is for 10 times overabundance of iron with $2\leq h\leq10$.
}
\label{fig:RMSfitting}
\end{figure}

Figure \ref{fig:RMSfitting} shows the observed rms spectrum of IRAS 13224--3809, in which results in the previous section are overplotted.
We use all the archival data of this object with {\it XMM-Newton} between 2002 and 2016, whose total exposure time is 1.46~Ms.
These data were reduced with the Science Analysis System (SAS) v15.0.0, the latest calibration files, and the standard threads of SAS.
We manually discarded high-background periods to get consecutive data as long as possible in each sequence.
A time-bin width of the rms spectra is 1500~s, which corresponds to the frequency investigated in the lag features ($[5.8-10.5]\times10^{-4}$~Hz).
The variability amplitude at 7 keV is 55\%, whereas that at 2 keV is 98\%.
In our calculations, when $h$ varies within $2.2-10$ (red dashed in figure \ref{fig:RMSfitting}), the rms dip is produced but the depth is too shallow to explain the observation.
We calculated the rms spectra with different $h$ ranges (green dotted and magenta dot-dashed in figure \ref{fig:RMSfitting}), but
none can explain the observed feature.
Therefore we conclude that the observed rms dip cannot be explained by the relativistic disc reflection model as far as the iron abundance is solar.

Next, we intentionally change the iron line flux to explain the deep rms dip.
We manually increase the iron flux by ten times, equivalently 10 solar overabundance of iron, and calculate the rms spectra.
As a result, the observed deep dip can be almost reproduced (blue solid).
Therefore, we see that the extreme iron overabundance ($\gtrsim10$ solar) is required to explain the rms dip with the relativistic light bending model.

\subsubsection{Lag features} \label{sec4.1.2}
\begin{figure}
\centering
\includegraphics[width=60mm,angle=270]{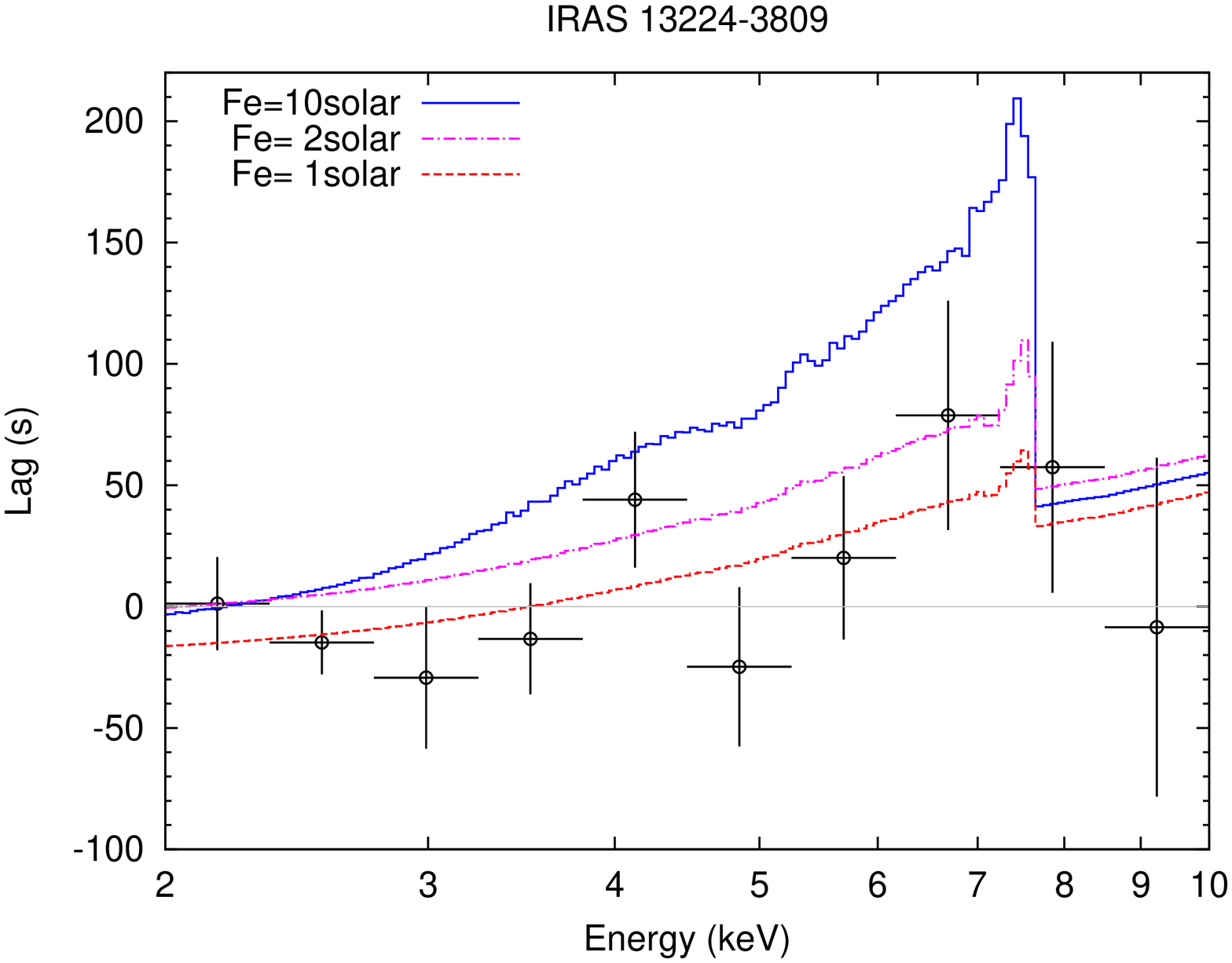}
\caption{
Observed lag-energy spectrum of IRAS 13224--3809 (black bins) in the frequency range of $[5.8-10.5]\times10^{-4}$~Hz,
compered with the relativistic disc reflection model with $a=0.998$.
The red dashed, magenta dot-dashed, and blue solid lines show when the iron abundance is equal to, twice, and 10 times of the solar abundance, respectively.
The zero point in both the data and model lines corresponds to the reference band of 2--10~keV.
}
\label{fig:compare_lamp}
\end{figure}

We calculate the lag-energy spectrum of IRAS 13224--3809 
in the frequency range of $[5.8-10.5]\times10^{-4}$~Hz (figure \ref{fig:compare_lamp}), 
following \citet{kar13c}.
We use the same data  in the rms spectrum.
The Fe-K lag amplitude is about 80~s, which corresponds to $\sim2-3\,R_g/c$.
Indeed, the lag-energy plot obtained by our simulation is consistent with the observed one (red dashed).
The lag amplitude becomes longer as the iron line flux (with time delay) is stronger.
The magenta line in figure \ref{fig:compare_lamp} shows the case when the iron flux is doubled,
which is still acceptable.
However, when the iron flux is ten times stronger (blue line), the lag amplitude gets more than 150~s, which is clearly rejected.
Thus, the iron abundance constrained by the lag feature ($\sim1-2$ solar) contradicts the one derived from the rms dip ($\gtrsim10$ solar).

\begin{figure}
\centering
\includegraphics[width=60mm,angle=270]{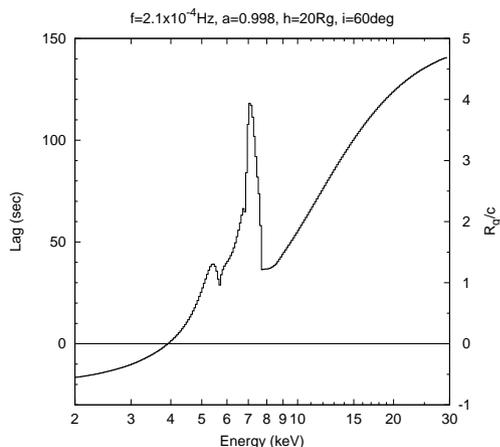}
\caption{
Lag-energy plots for $h=20\,R_g$ and $a=0.998$ at the frequency of $2.1\times10^{-4}$~Hz. 
}
\label{fig:lagE2}
\end{figure}

\citet{kar13c} reported flux-dependence of the Fe-K lags in this object.
They examined lags from low- and high-flux intervals, where average count rates are different by one order of magnitudes.
The Fe-K lags disappear in the high-flux period for $[1.4-2.8]\times 10^{-4}$~Hz, whereas
they are seen in the low-flux period for $[5.8-10.5]\times 10^{-4}$~Hz (figure 6 in \citealt{kar13c}).
In the framework of the relativistic light bending model,
this change may be explained by significant change of the lamp post height.
As the lamp post height varies, the RDC flux is hardly variable and the PLC flux varies by an order of magnitudes, therefore
the lag in the high-flux period is diluted by the strong PLC component.
However, our calculation indicates that 
the reverberation lag is still clearly seen even when $h=20$ (figure \ref{fig:lagE2}).
Assuming that the frequency where the lag amplitude gets to zero monotonically decreases with larger $h$, 
the height where the lag disappears should be $\sim40\,R_g$.
However, this clearly contradicts the observed line profile;
the line is hardly skewed and broadened with such a large source height (e.g., \citealt{fab02}).

\subsection{On the black-hole spin constraint}
\begin{figure}
\centering
\includegraphics[width=60mm,angle=270]{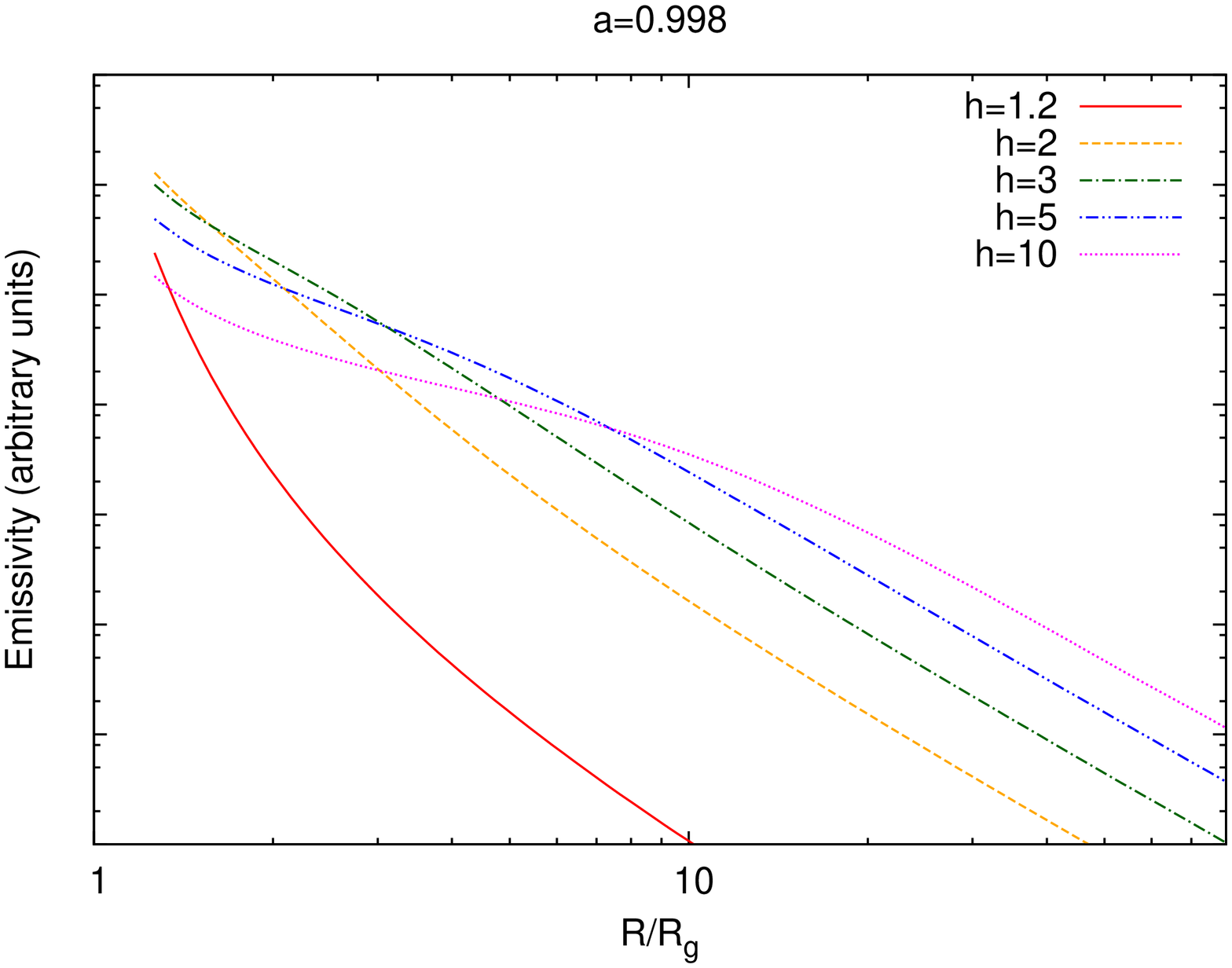}\\
\includegraphics[width=60mm,angle=270]{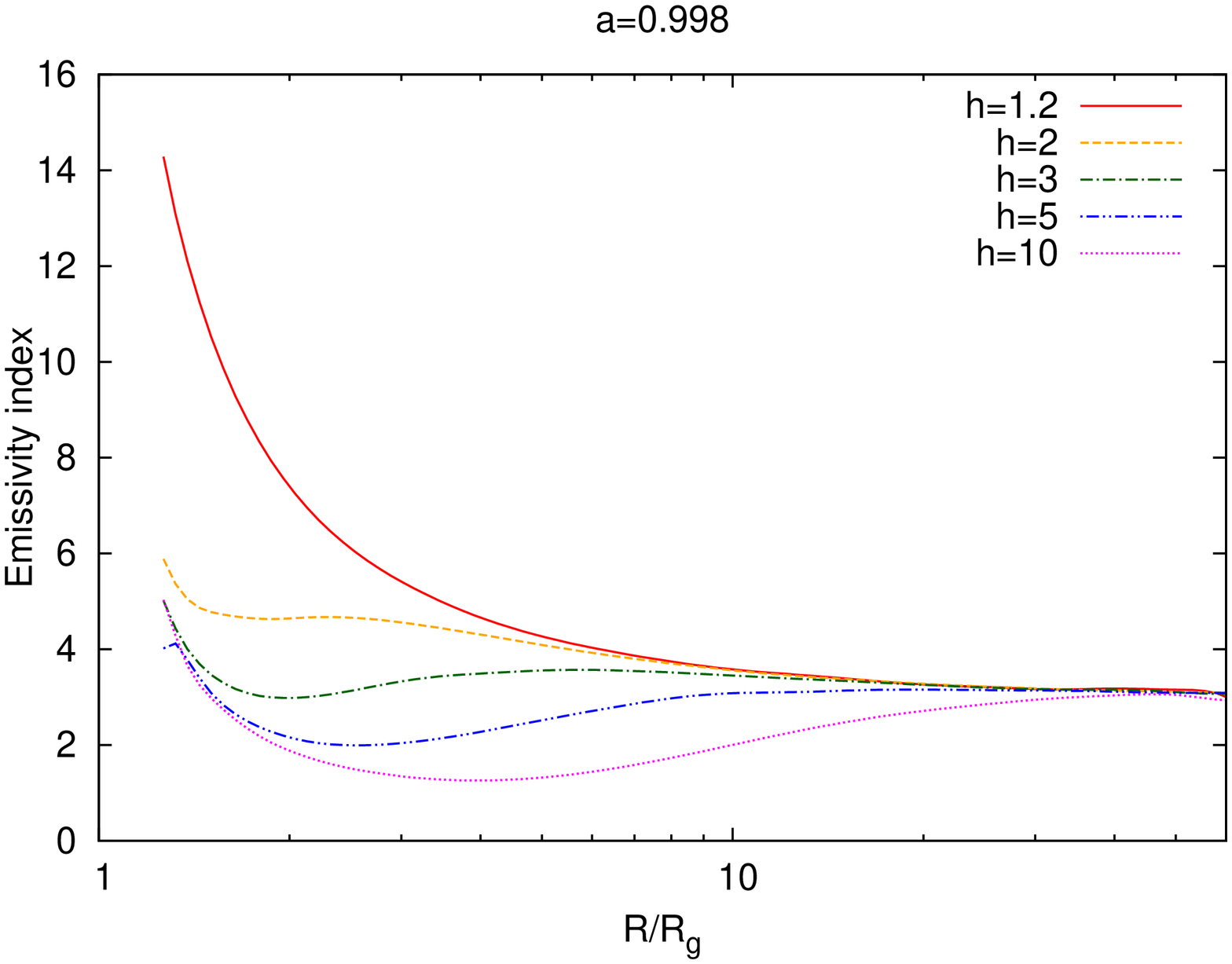}
\caption{
(Upper) Disc emissivity profiles, $\epsilon(r)$, calculated by our ray-tracing method for different lamp-post heights.
(Lower) Emissivity indices, $q$, which is defined by $\epsilon(r)\propto r^{-q}$.
}
\label{fig:emissivity}
\end{figure}

Some authors claimed that spin of IRAS 13224--3809 is tightly constrained to be almost maximum based on model fitting of the X-ray spectra, e.g., $0.990_{-0.003}^{+0.001}$ and $0.988\pm0.001$ (\citealt{fab13}; \citealt{chi15}).
They proposed extremely-high emissivity indices $q\,(\gtrsim9)$,
where the disc emissivity per area at a radius $r$ is proportional to $r^{-q}$.
When the emissivity index is so high,
the reflection component from the inner region of the disc is fully dominant, and thus
the Fe-K line profiles are extremely sensitive to small change of $a$ (i.e.\ change of $r_{\rm ms}$).
We calculate the emissivity profiles and indices from our simulations (figure \ref{fig:emissivity}).
In this figure, the emissivity index is up to $\sim6$ at the innermost edge of the disc with the source height of $2\,R_g$ even in the maximum spin case, and
such a high emissivity index of $q\gtrsim9$ is possible only when the lamp locates within $2\,R_g$ from the central BH.
Thus, the spin parameters cannot be constrained 
unless the source height is extremely small (see also \citealt{bon16}).
However, our model requires significant height variations (typically $h=2-10$) to produce the observed Fe-K rms dip (figure \ref{fig:RMSfitting}).
Also, we might tell whether $a<0.9$ or $a>0.9$ from the time variability, but
it is difficult to distinguish $a=0.9$ from $a=0.998$ (figures \ref{fig:rms}, \ref{fig:lagE}).

The situation becomes more complex since other mechanisms also affect the Fe-K spectral feature.
\citet{nod11a} pointed out that the iron line profiles are easily changed when the continuum level is different.
In addition, absorption features can mimic the broad emission line;
when the cold clumpy absorbers partially cover the X-ray source, a broad Fe-K absorption edge is produced  (e.g., \citealt{tan04,miy12,miz14,iso16,yam16}), and
when the strong ultrafast outflow exists in the line of sight (e.g., \citealt{tom10}), the broad and blueshifted absorption line features due to H- or He-like iron ion are seen in the energy spectra \citep{hag15,hag16}.
These absorption features affect the iron line profile, and thus
we suggest that the BH spin values estimated from Fe-line spectral fitting or time variations are not reliable.

%%%%%

%%%%%%%%%%%%%%%%%%%%%%%%%%%%%%%%%%%%%%%%%%%%%%%%%%%%%%%%%%%%%%%%%%%%%%%%%%%%
\section{Conclusion} \label{sec5}

Seyfert galaxies are known to exhibit characteristic spectral variability; root-mean-square (rms) dips in the Fe-K energy band and the Fe-K reverberation lags.
The relativistic light bending model, where a lamp post is moving vertically along the rotation axis,
has been proposed to explain these features.
We critically examined whether the relativistic light bending model 
can explain the two features simultaneously.
A ray-tracing technique is adopted including the full treatment of general relativity to calculate photon paths and delay times of the reflection.
We try to explain Fe-K features of IRAS 13224--3809, which is a representative source to exhibit the characteristic Fe-K features.
The observed rms dip may be reproduced when an extreme iron overabundance ($\gtrsim10$ solar) is assumed.
On the contrary, the observed lag-energy spectrum requires solar abundance of  iron;
it clearly rejects the extreme iron overabundance.
In addition, the observed diminution of the lag feature in the high-flux state
needs a very large source height of $\sim40\,R_g$,
which contradicts the observed broad line profile.
Consequently, we conclude that the relativistic light bending model cannot explain both the rms dips and reverberation lags in the Fe-K band simultaneously.
BH spin parameters constrained by this model are thus not reliable.

\begin{ack}
 Authors are financially supported by the JSPS/MEXT KAKENHI Grant Numbers JP15J07567 (MM), JP17J08829 (KM), JP16K05309 (KE), and JP24105007, JP15H03642 (MT).
NK is supported by the Hakubi project in Kyoto University.
\end{ack}

\bibliographystyle{aa}
\bibliography{00}
%%%%%%%%%%%%%%%%%%%%%%%%%%%%%%%%%%%%%%%%%%%%%%%%%%

%%%%%%%%%%%%%%%%% APPENDICES %%%%%%%%%%%%%%%%%%%%%

\appendix
%%%%%%%%%%%%%%%%%%%%%%%%%%%%%%%%%%%%%%%%%%%%%%%%%%
\section{Numerical calculation of the relativistic light bending model} \label{app1}
We explain details of the relativistic light bending model and the numerical calculation methods used in this paper.
In this section, we use ``geometrised units'' in which $G$ and $c$ are set to be unity.

\subsection{Relativistic light bending model}\label{app1:1}

%% Primary photons
First, we consider the photons to directly reach the observer.
The height, radial position and 4-velocity of the source are denoted by $h_{\rm s}$, $\rho_{\rm s}$ and $u^{\mu}_{\ell}=(u^{0}_{\ell},0,0,u^{3}_{\ell})$ (see figure \ref{abs}).
In its inertial frame, the radiation energy per unit time, and energy, $d W_{\ell}(e_{\ell}, t)/dt$, is defined as
\begin{equation}
\frac{d W_{\ell}(e_{\ell}, t)}{dt}de_{\ell}=N e_{\ell}^{-\Gamma+1}\delta(t)de_{\ell},
\end{equation}
where $N$ is the photon number, $e_{\ell}$ is the photon energy, $t$ is the Boyer-Lindquist time coordinate, 
and $\Gamma\ (=2)$ is the spectral index.
The primary continuum emission to reach directly to the observer from the emitter 
is expressed as
\begin{eqnarray}
\frac{dW_{\rm p}(e_{\rm obs},t)}{dt}de_{\rm obs}
&=&g_{0}^2 \frac{dW_{\ell}(e_{\ell},t-\tau_{\rm p})}{dt}de_{\ell} \nonumber\\
&=& g_0^{\Gamma}N e_{\rm obs}^{-\Gamma+1}\delta(t-\tau_{\rm p})de_{\rm obs},
\end{eqnarray}
where $e_{\rm obs}$ is the observed photon energy, and $\tau_{\rm p}(=\ell_{\rm p})$ is 
the time taken for a photon from the source to reach the observer's plane.
Further, $g_0$ is the energy-shift factor between the observer and source, and is written as
\begin{equation}
g_0 = \frac{e_{\rm obs}}{e_{\ell}} 
= \frac{1}{u_{\ell}^{0}(1-\Omega_{\ell}\Lambda)},
\end{equation}
where $\Omega_{\ell}=u^{3}_{\ell}/u^{0}_{\ell}$ is the angular velocity of the source, and 
$\Lambda$ is the angular momentum of the photon with respect to the azimuth direction ($\phi$) per energy.
Therefore, the total photon flux to reach the observer, $F_{\rm p}(e_{\rm obs},t)de_{\rm obs}$, is written as
\begin{equation}
F_{\rm p}(e_{\rm obs},t)de_{\rm obs}\label{pri}
= g_0^{\Gamma} N'_{\rm p} e_{\rm obs}^{-\Gamma}\delta(t-\tau_{\rm p})de_{\rm obs}, 
\end{equation}
where $ N'_{\rm p}$ is the photon number to reach the observer's plane per unit area.

%%%%%
%% Incident photon to disc
Next, we consider the incident photon in the comoving frame of the accretion disc.
The incident energy flux entering the gas element of 
the disc located at $(r,\phi)$, is expressed as
\begin{eqnarray}
\frac{dW_{\rm d}(e_{\rm d},t)}{dtdS_{\rm d}}de_{\rm d}
&=&  g_{1}^2\frac{dW_{\ell}(e_{\ell}, t-\tau_{\rm d})}{dtdS_{\rm d}}de_{\ell} \nonumber\\
&=&  g_{1}^{\Gamma}N'_{\rm d} e_{\rm d}^{-\Gamma+1}\delta(t-\tau_{\rm d})de_{\rm d},
\end{eqnarray}
where $N'_{\rm d}$ is the incident photon number per unit area, $dS_{\rm d}$ is the disc area element, 
$e_{\rm d}$ is the energy of the incident photon in the comoving frame, and
$\tau_{\rm d}$ is the photon path length between the source and gas element  (see figure \ref{abs}).
Further, the energy shift factor between the source and disc, $g_1$, is written by
\begin{equation}
g_{1}
= \frac{u_{\rm d}^{0}(1-\Omega_{\rm d}\Lambda)}{u_{\ell}^0(1 -\Omega_{\ell} \Lambda)},
\end{equation}
where $u_{\rm d}^{\mu}$ and $\Omega_{d}$ is the 4-velocity and angular velocity of the disc element.

%% re-emission from the disc
Photons in the fluorescent K$\alpha$ line and the reflected continuum from the disc element are regarded as the reflection component.
Probabilities that the fluorescent $\rm K\alpha$ line (6.4~keV at the rest frame) and the reflected continuum are emitted from the disc element, $P_{\rm K\alpha}$ and  $P_{\rm scat}$, are expressed as
\begin{eqnarray}
P_{\rm K\alpha}(e_{\rm d}) &=& W_{\rm K}\frac{\sigma_{\rm a}(e_{\rm d})}{\sigma_{\rm a}(e_{\rm d})+\sigma_{\rm s}(e_{\rm d})}, \\
P_{\rm scat}(e_{\rm d})   &=& \frac{\sigma_{\rm s}(e_{\rm d})}{\sigma_{\rm a}(e_{\rm d})+\sigma_{\rm s}(e_{\rm d})},
\end{eqnarray}
where $e_{\rm d}$ is the energy of the incident photon in the comoving frame,
$W_{\rm K}[=0.35\ (0)\ {\rm for}\ e\geq E_{\rm e}=7.112\,{\rm keV}\ (e<E_{\rm e})]$ 
is the iron K$\alpha$ fluorescence yield \citep{pio92}, 
$\sigma_{\rm a}(e)$ is the iron photoelectric absorption cross-section 
(table 2 in \citealt{mor83}), and $\sigma_{\rm s}(e)$ is the electron
scattering cross-section given by the Klein-Nishina formula, such as
\begin{equation}
\sigma_{\rm a}(e)=\sigma_{\rm T}\frac{3}{4}\left[
\frac{1+e}{e^3}\left\{
\frac{2e(1+e)}{1+2e}-\ln(1+2e)
\right\}+\frac{\ln(1+2e)}{2e}-\frac{1+3e}{(1+2e)^2}\right],
\end{equation}
where $\sigma_{\rm T}$ is the Thomson cross-section.

In the comoving frame of the gas element at $(r,\phi)$, 
reflected photon flux, $F_{\rm d}(e_{\rm d},t)$ is expressed as
\begin{equation}
F_{\rm d}(e_{\rm d},t)= F_{\rm K\alpha}(e_{\rm d},t)+F_{\rm scat}(e_{\rm d},t),
\end{equation}
where $F_{\rm K\alpha}$ and $F_{\rm scat}$ are the photon flux of 
the fluorescent K$\alpha$ line and reflected continuum radiation, and are defined as 
\begin{equation}
\begin{array}{ll}
F_{\rm K\alpha}(e_{\rm d},t)
&= g_1^{\Gamma}N'_{d}\delta(e_{\rm d}-6.40\,{\rm keV})\delta (t-\tau_{\rm d})
\int^{\infty}_{E_{\rm e}} \! P_{\rm K\alpha}(e)e^{-\Gamma}de\\
F_{\rm scat}(e_{\rm d},t)
&= g_{1}^{\Gamma}N'_{\rm d}\delta (t-\tau_{\rm d})e_{\rm d}^{-\Gamma}P_{\rm scat}(e_{\rm d}). \label{line}
\end{array}
\end{equation}
Therefore the radiative energy per unit time, energy, and solid angle is written as
\begin{equation}
\frac{dW_{\rm rev}(e_{\rm d},t)}{dtd\Omega dS_{\rm d}} 
= \frac{e_{\rm d}}{4\pi}F_{\rm d}(e_{\rm d},t).
\end{equation}
The emissivity of the fluorescent K$\alpha$ line, $j$, is proportional to $F_{\rm K\alpha}$:
\begin{equation}
j(r,\phi)=J  g_1(r,\phi)^{\Gamma}N'_{d}(r,\phi)\int^{\infty}_{E_{\rm e}} P_{\rm K\alpha}(e)e^{-\Gamma}de \ \ \ (J={\rm const.}).
\end{equation}

%% scattering photon to observer
Finally, we consider the total radiation energy of the photons 
to reach the observer.
The observed radiation emitted from a disc element located at $(r,\phi)$ is expressed as
\begin{eqnarray}
\left[\frac{dW_{\rm obs}(e_{\rm obs},t)}{dtd\Omega dS_{\rm obs}}\right]_{j}
&=& \left[g_{2}^2 \frac{e_{\rm obs}}{4\pi}F_{\rm d}(e_{\rm obs}/g_2,t-\tau_{\rm r})\right]_{j} \label{grid}\\ 
g_{2}&=& \frac{e_{\rm obs}}{e_{\rm d}}=\frac{1}{u^0 (1-\Omega_{\rm d}\Lambda_{\rm r})} \\
\Lambda_{\rm r} &=&-x_{\rm obs} \sin \theta_{\rm obs},
\end{eqnarray}
where $(dS_{\rm obs})_{j}$ is the disc element, 
$e_{\rm obs}$ is the observed photon energy, 
$\tau_{\rm r}$ is the time taken for a photon
from the gas element to reach the observer's plane.
The ray reaches the Cartesian coordinate on the observer's plane, $(x_{\rm obs}, y_{\rm obs})$,
with the angular momentum of the photon with respect to the $\phi$ direction per energy, 
$\Lambda_{\rm r}$ (figure \ref{abs}).
The suffix "j" is the number of each cell of the observer's plane.

Thus the total energy flux to reach the observer's plane is expressed as
\begin{eqnarray}\label{total}
F_{\rm tot}&=&\sum_{j}\left( \frac{dW_{\rm obs}(e_{\rm obs},t-\tau_{\rm r})}{dt dS_{\rm obs}}de_{\rm obs}\right)_{j} \nonumber \\
&=& \sum_{j} \left[\frac{dS_{\rm obs}}{4\pi D^2} g_{2}^2 F_{\rm d}(e_{\rm obs}/g_2,t-\tau_{\rm r})\right]_{j}.
\end{eqnarray}

\subsection{Methods of numerical calculation}\label{numerical}

We numerically calculate trajectories of the rays emitted from
the static source by solving the null geodesic equation;
we apply the symplectic method \citep{yos93,mor15}, 
where the impact parameter of each ray is given by \citet{kar92}.
In this calculation, we obtain $\tau_{\rm p}$, $N'_{\rm p}$ and $g_{0}$, and then 
calculate $F_{\rm p}$ by using equation (\ref{pri}).
Next, we calculate trajectories to reach the disc surface from the source, and obtain $\tau_{\rm d}$, $N'_{\rm d}$, $i$, and $g_{1}$.
Then we calculate $F_{\rm K\alpha}$ and $F_{\rm scat}$ by using equation (\ref{line}).

We calculate ray trajectories which leave each area element
of the observer's plane in the perpendicular direction to reach 
the disc by applying the symplectic method to obtain 
$\tau_{\rm r}(x_{\rm obs}, y_{\rm obs})$, and $g_2$.
$F_{\rm tot}$ is calculated by substituting $F_{\rm K\alpha}, F_{\rm scat}$, $\tau_{\rm r}(x_{\rm obs}, y_{\rm obs})$, and $g_2$ in equation (\ref{total}).

%%%%%%%%%%%%%%%%%%%%%%%%%%%%%%%%%%%%%%%%%%%%%%%%%%
\section{Simulation in the low inclination case} \label{app2}
Resultant rms spectra and lag features for $i=60$~deg are shown in the main text.
In this section, we explain those for the low inclination case, $i=30$~deg.

\begin{figure*}
\centering
\subfigure{
\resizebox{8cm}{!}{\includegraphics[width=60mm,angle=270]{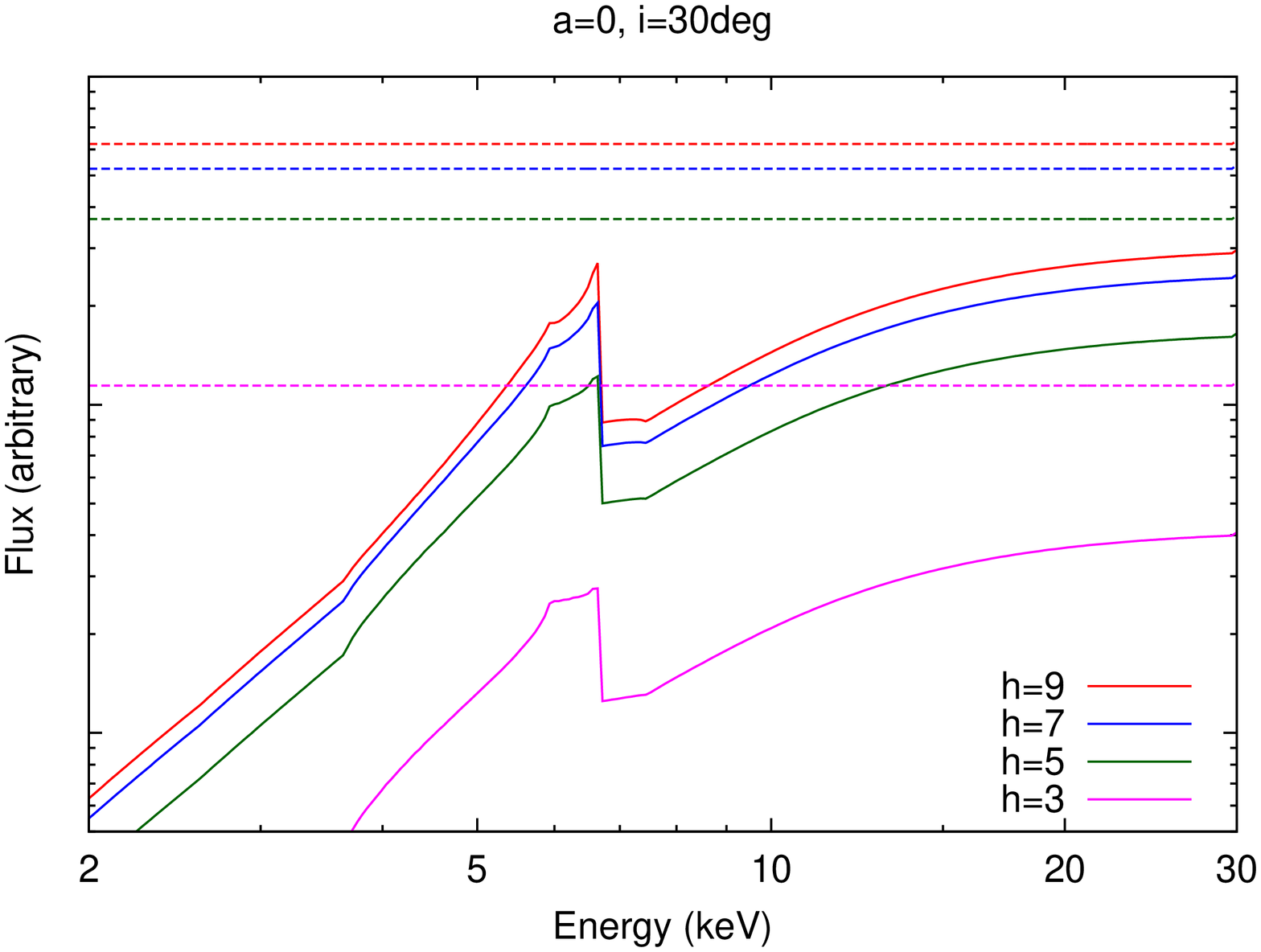}}
\resizebox{8cm}{!}{\includegraphics[width=60mm,angle=270]{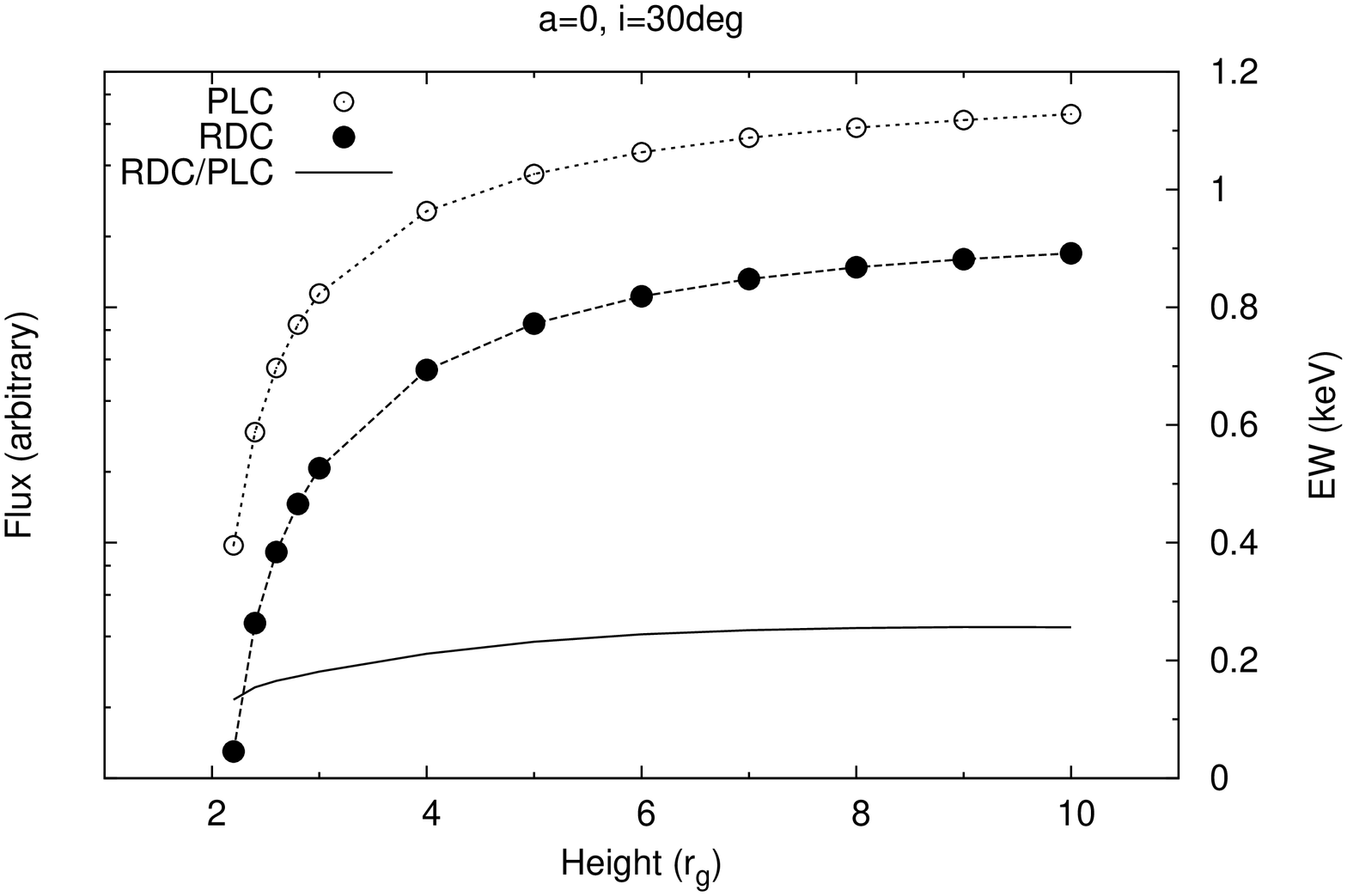}}
}
\subfigure{
\resizebox{8cm}{!}{\includegraphics[width=60mm,angle=270]{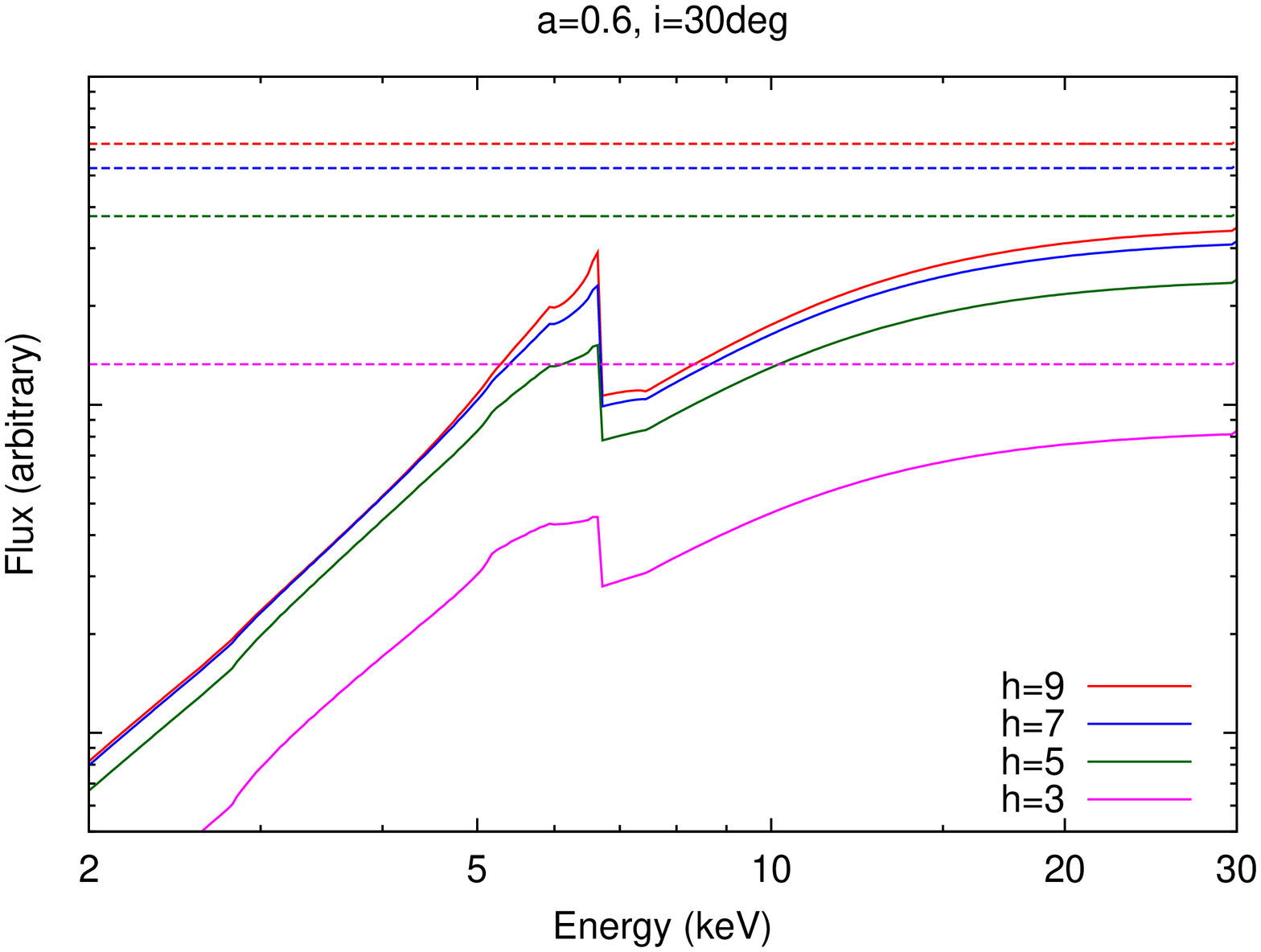}}
\resizebox{8cm}{!}{\includegraphics[width=60mm,angle=270]{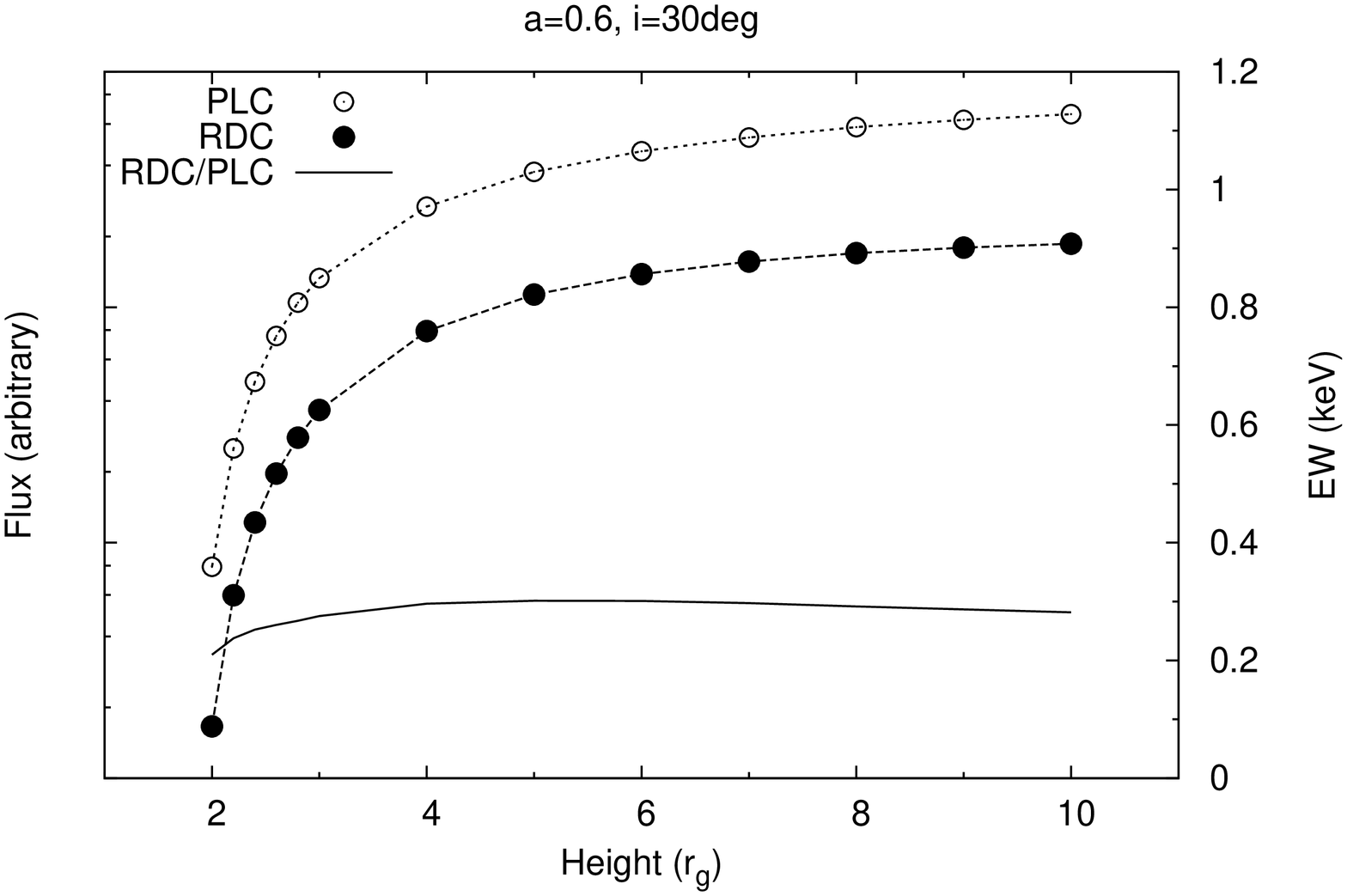}}
}
\subfigure{
\resizebox{8cm}{!}{\includegraphics[width=60mm,angle=270]{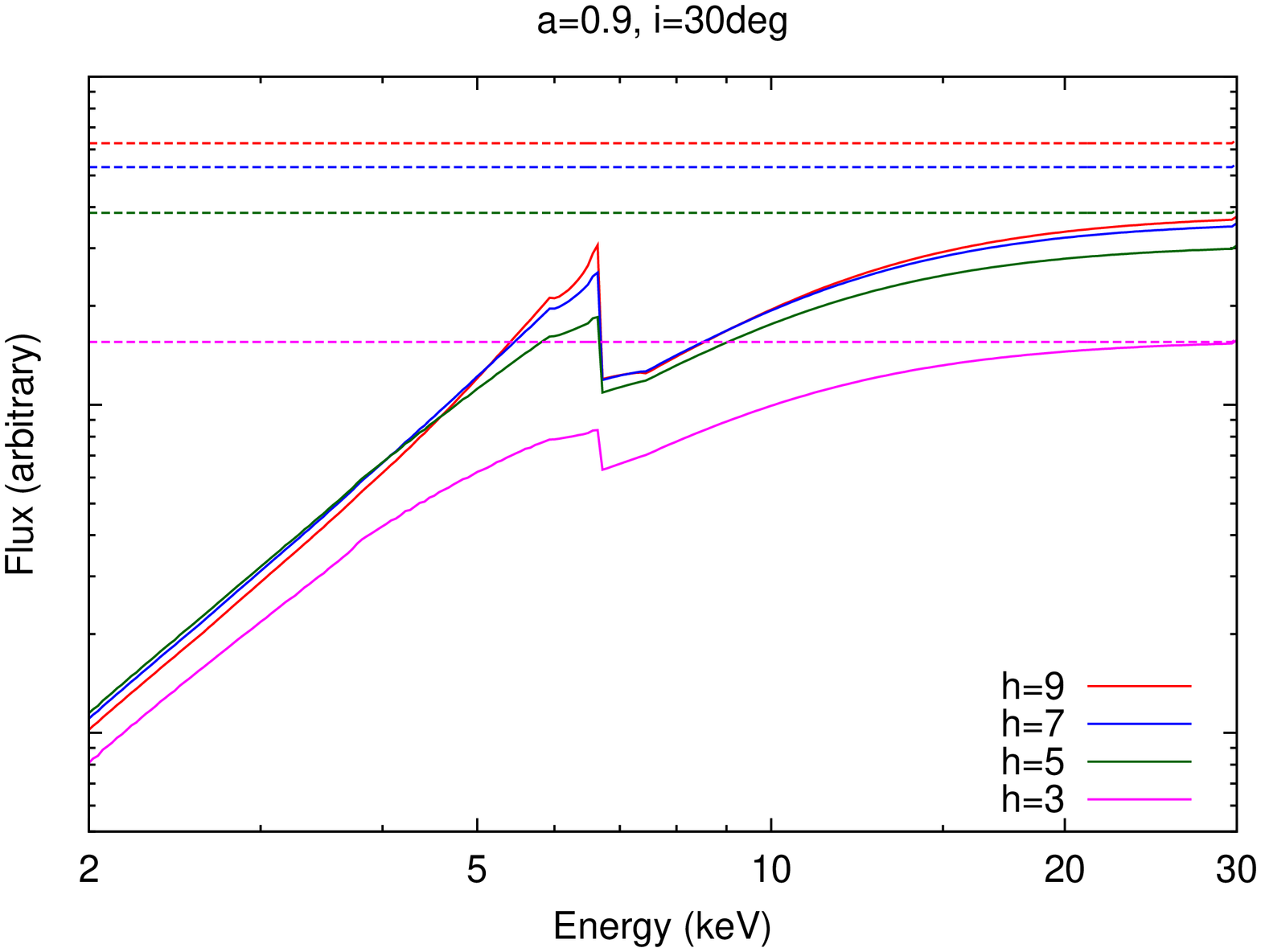}}
\resizebox{8cm}{!}{\includegraphics[width=60mm,angle=270]{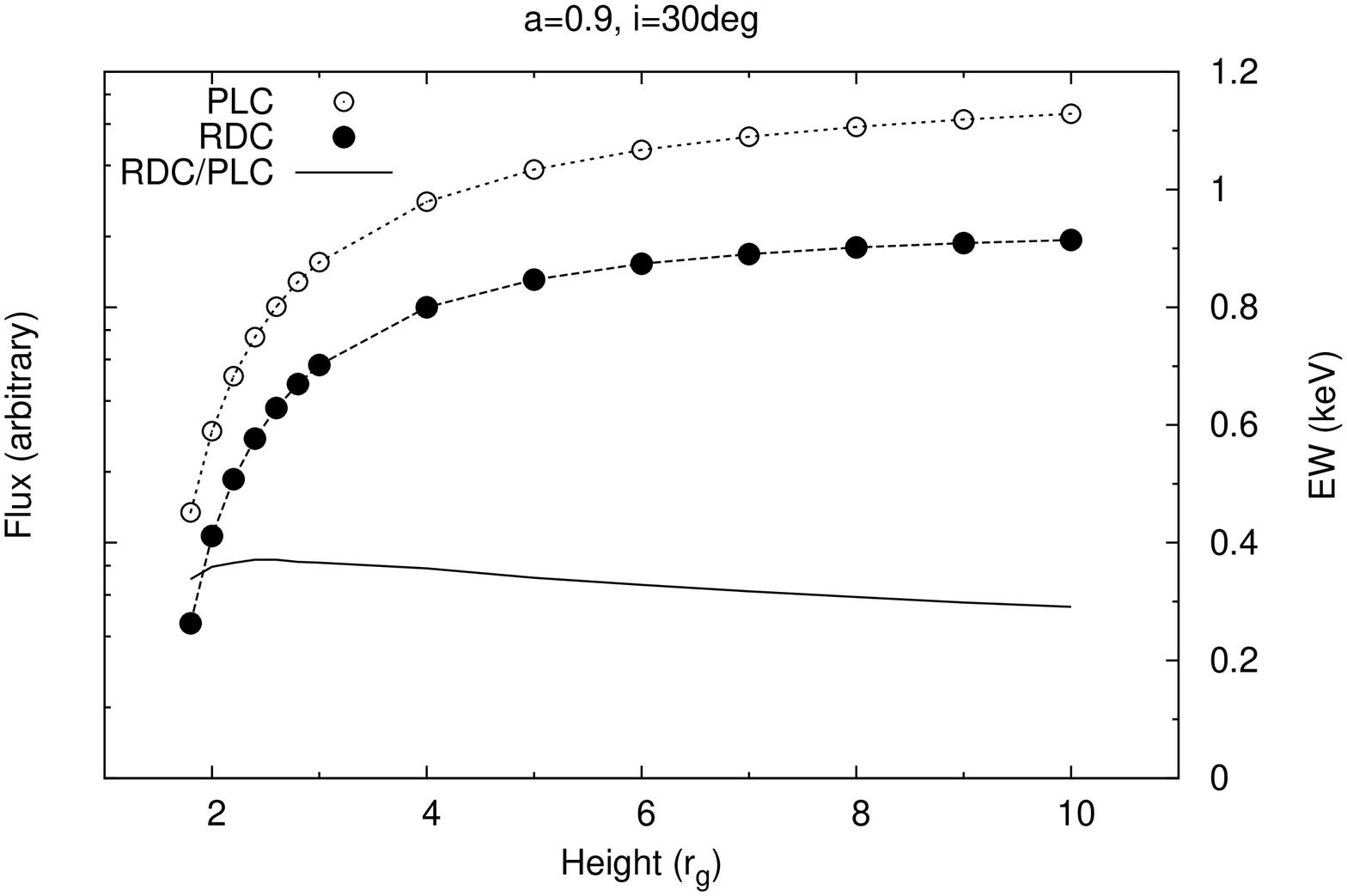}}
}
\subfigure{
\resizebox{8cm}{!}{\includegraphics[width=60mm,angle=270]{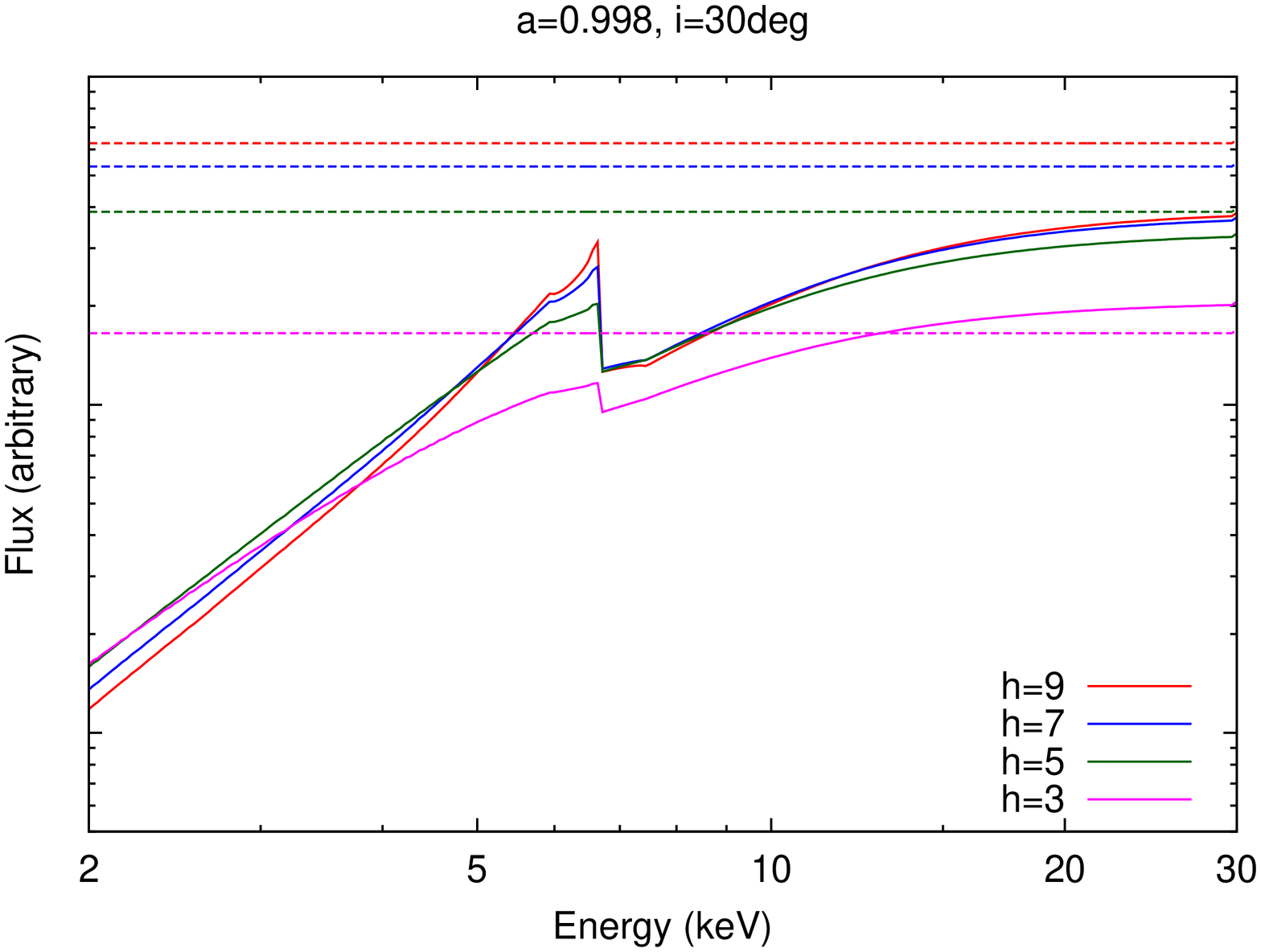}}
\resizebox{8cm}{!}{\includegraphics[width=60mm,angle=270]{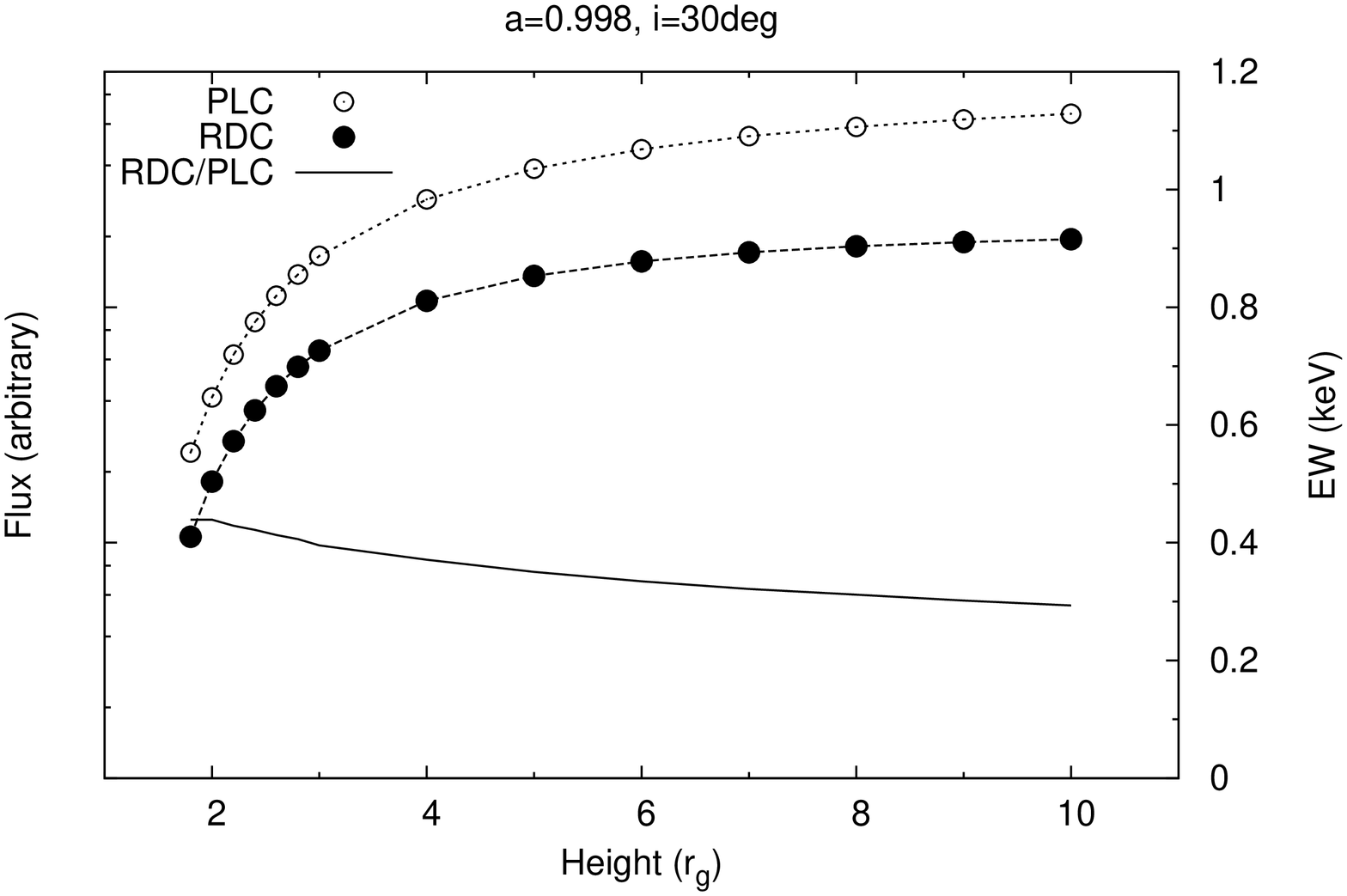}}
}
\caption{
Same as figure \ref{fig:ironspectra}, but for $i=30$~deg.
}
\label{fig:ironspectra2}
\end{figure*}

\begin{figure*}
\begin{center}
\subfigure{
\resizebox{8cm}{!}{\includegraphics[width=60mm,angle=270]{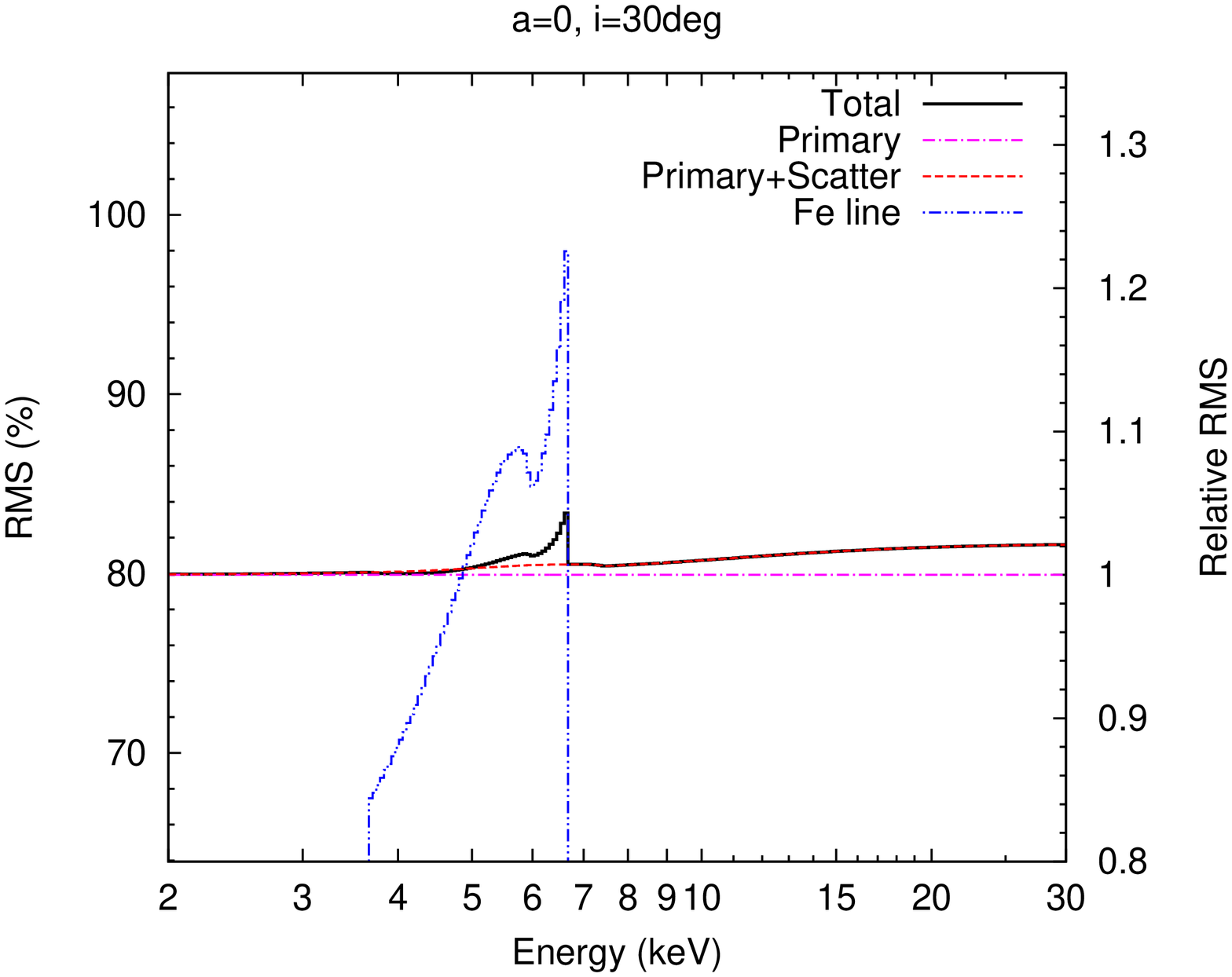}}
\resizebox{8cm}{!}{\includegraphics[width=60mm,angle=270]{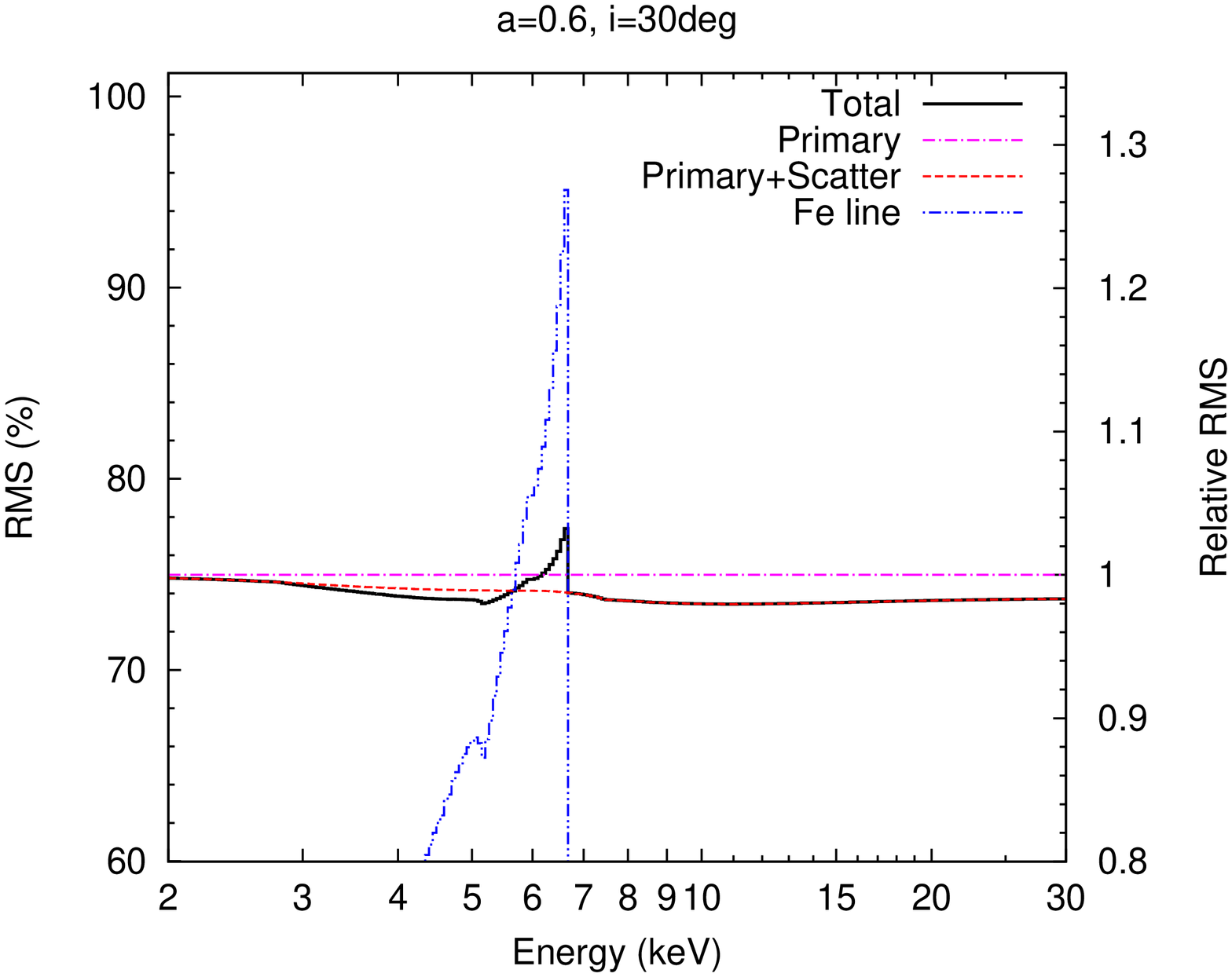}}
}
\subfigure{
\resizebox{8cm}{!}{\includegraphics[width=60mm,angle=270]{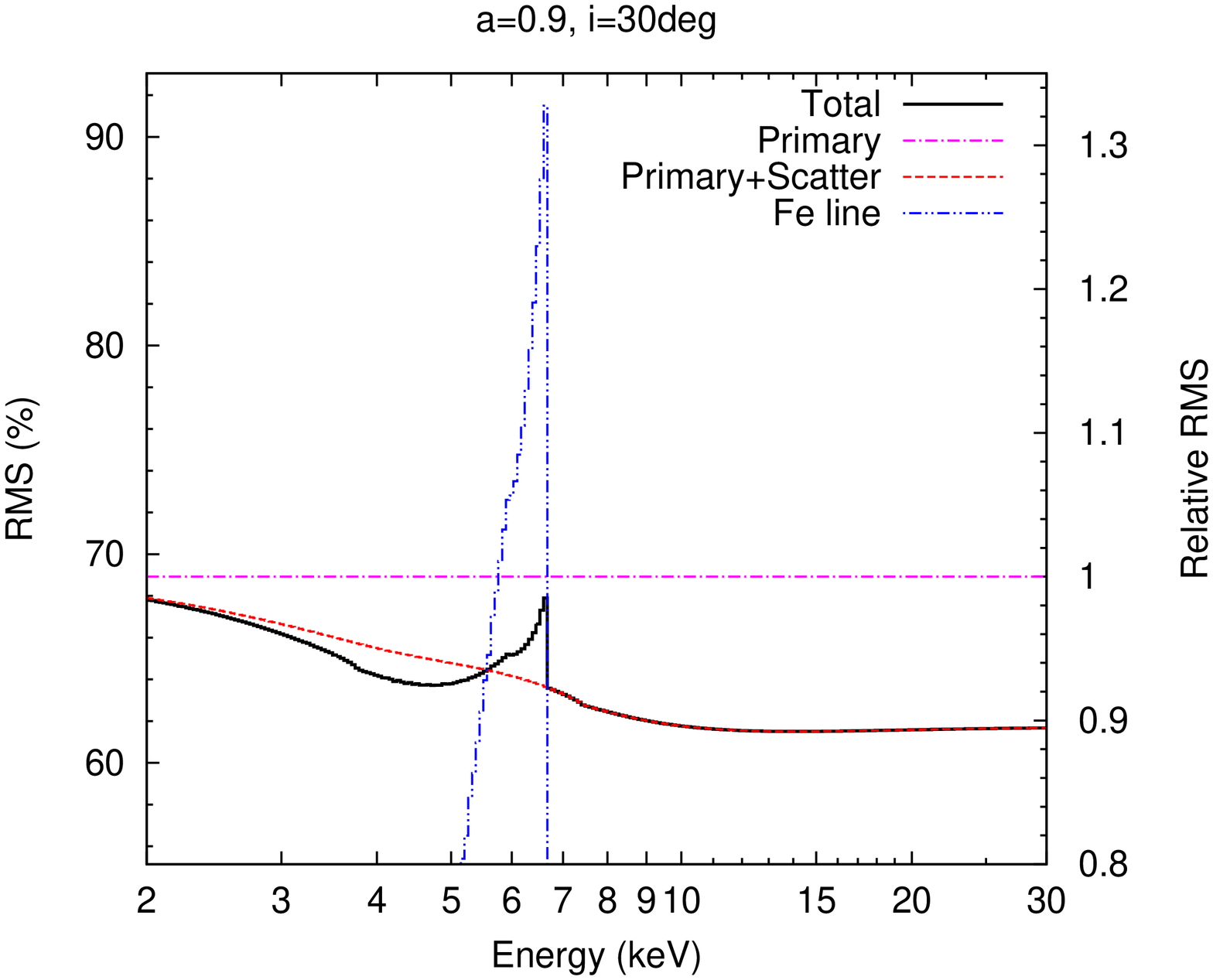}}
\resizebox{8cm}{!}{\includegraphics[width=60mm,angle=270]{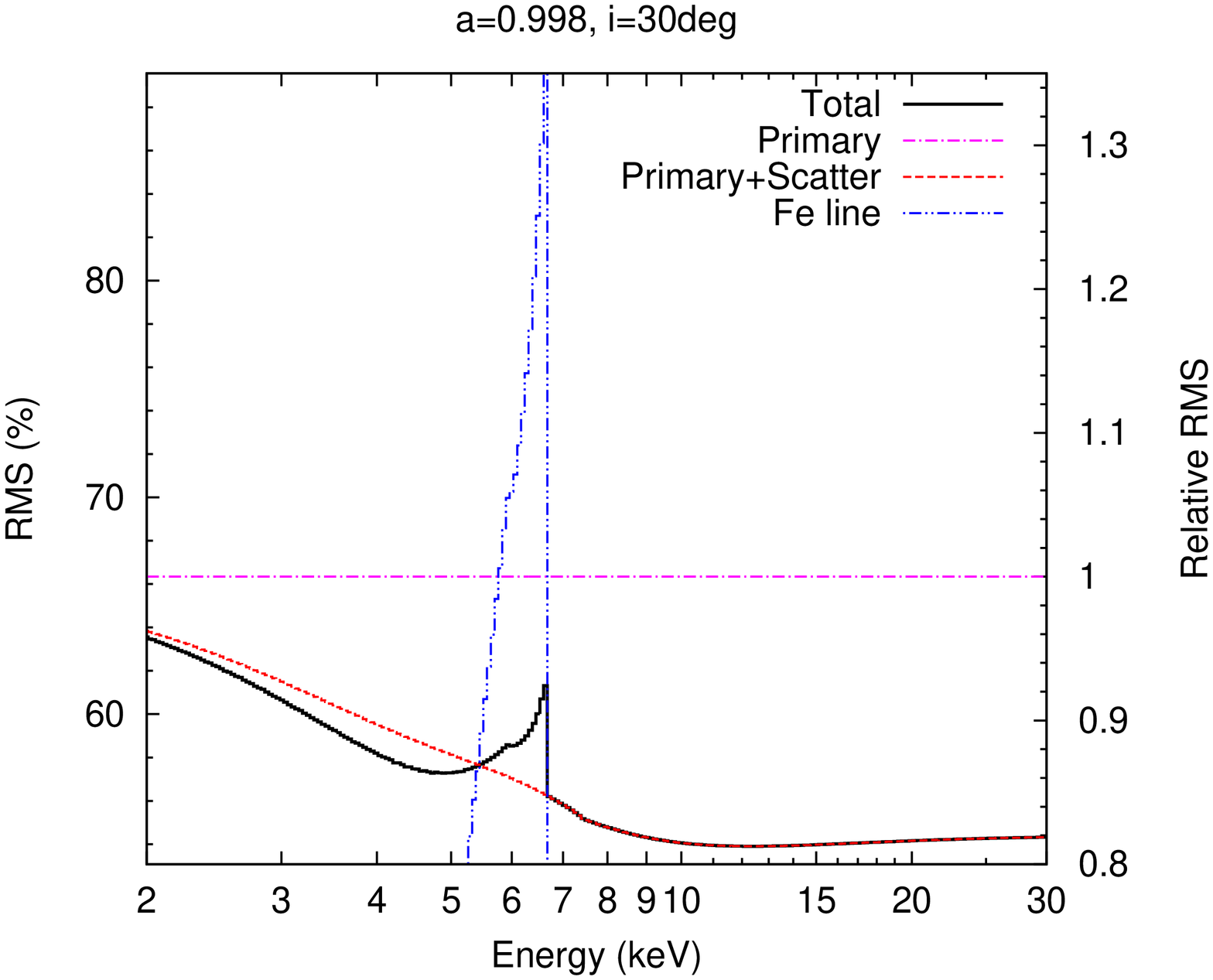}}
}
\caption{
Same as figure \ref{fig:rms}, but for $i=30$~deg
}
\label{fig:rms_30}
\end{center}
\end{figure*}

\begin{figure*}
\begin{center}
\includegraphics[width=150mm]{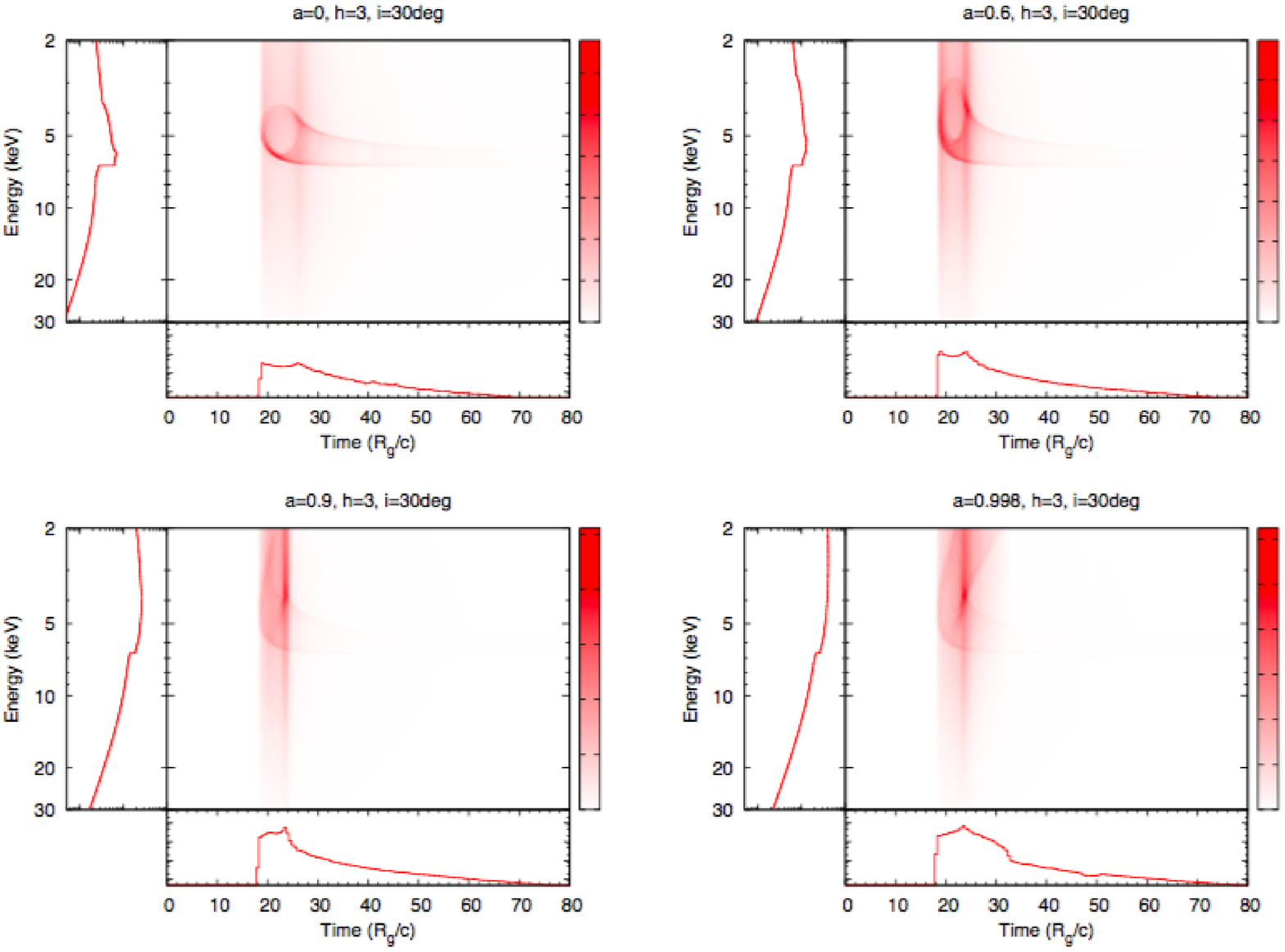}
\caption{
Same as figure \ref{fig:2dtf}, but for $i=30$~deg
}
\label{fig:2dtf_30}
\end{center}
\end{figure*}

\begin{figure*}
\begin{center}
\subfigure{
\resizebox{8cm}{!}{\includegraphics[width=65mm,angle=270]{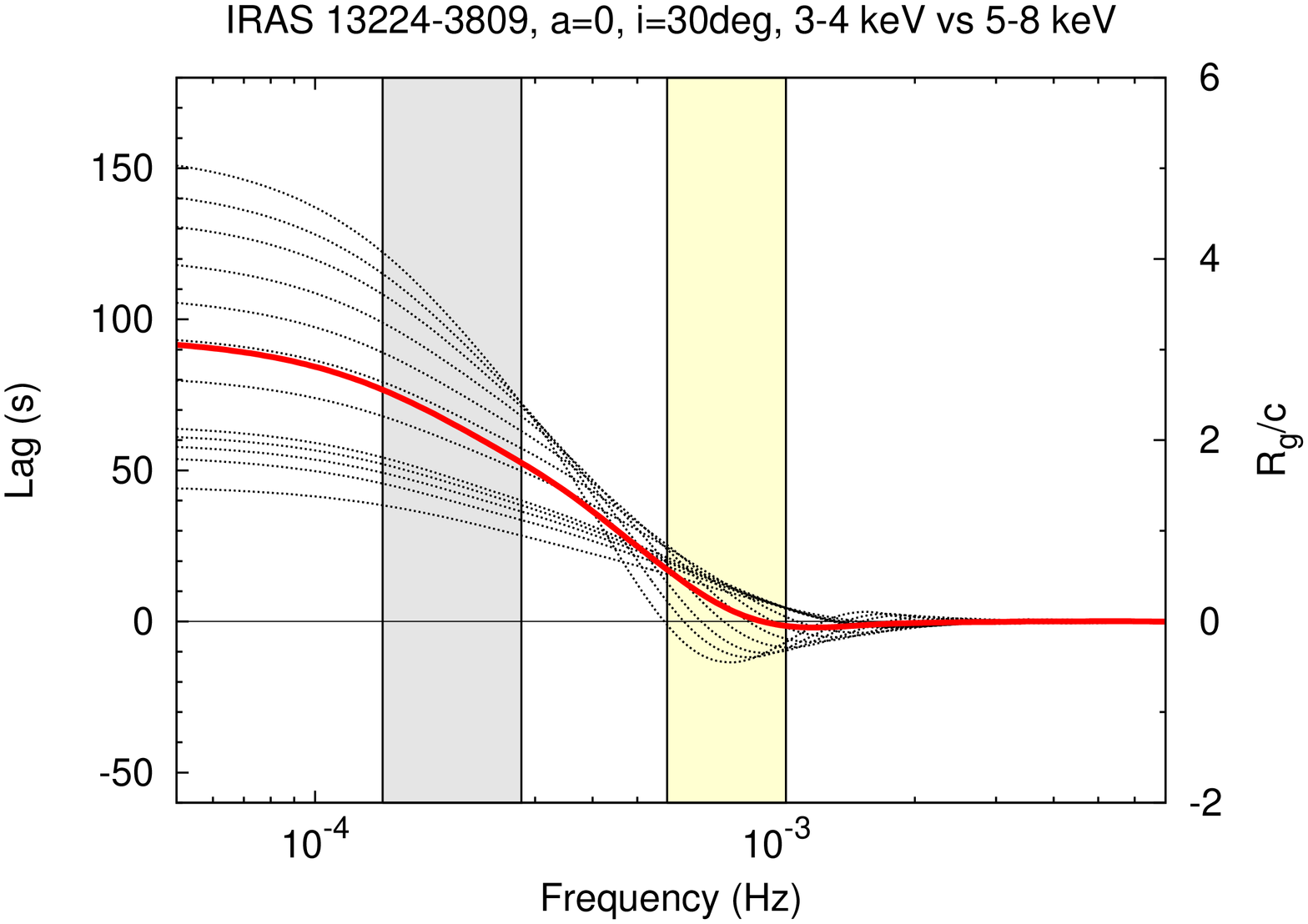}}
\resizebox{8cm}{!}{\includegraphics[width=65mm,angle=270]{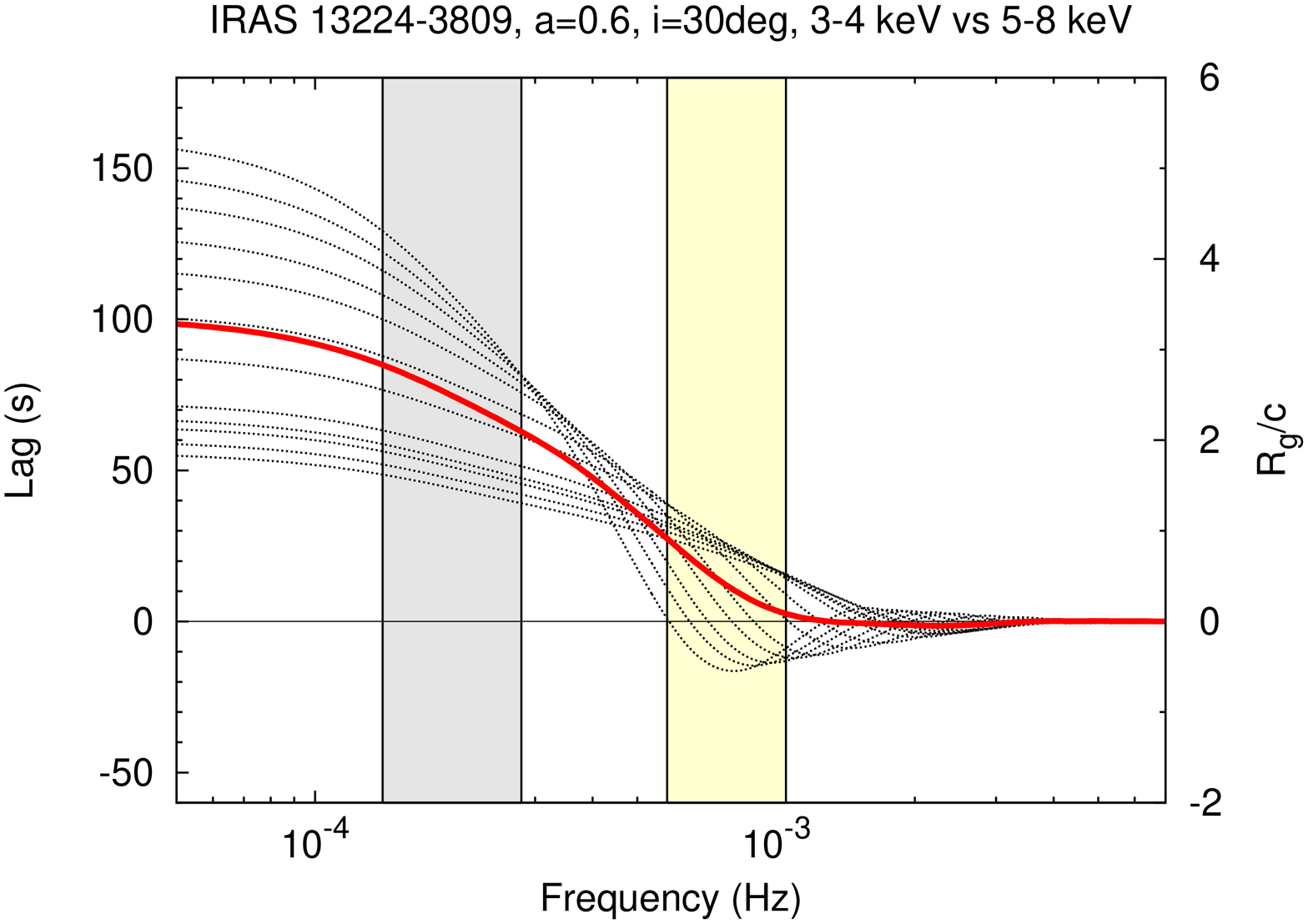}}
}
\subfigure{
\resizebox{8cm}{!}{\includegraphics[width=65mm,angle=270]{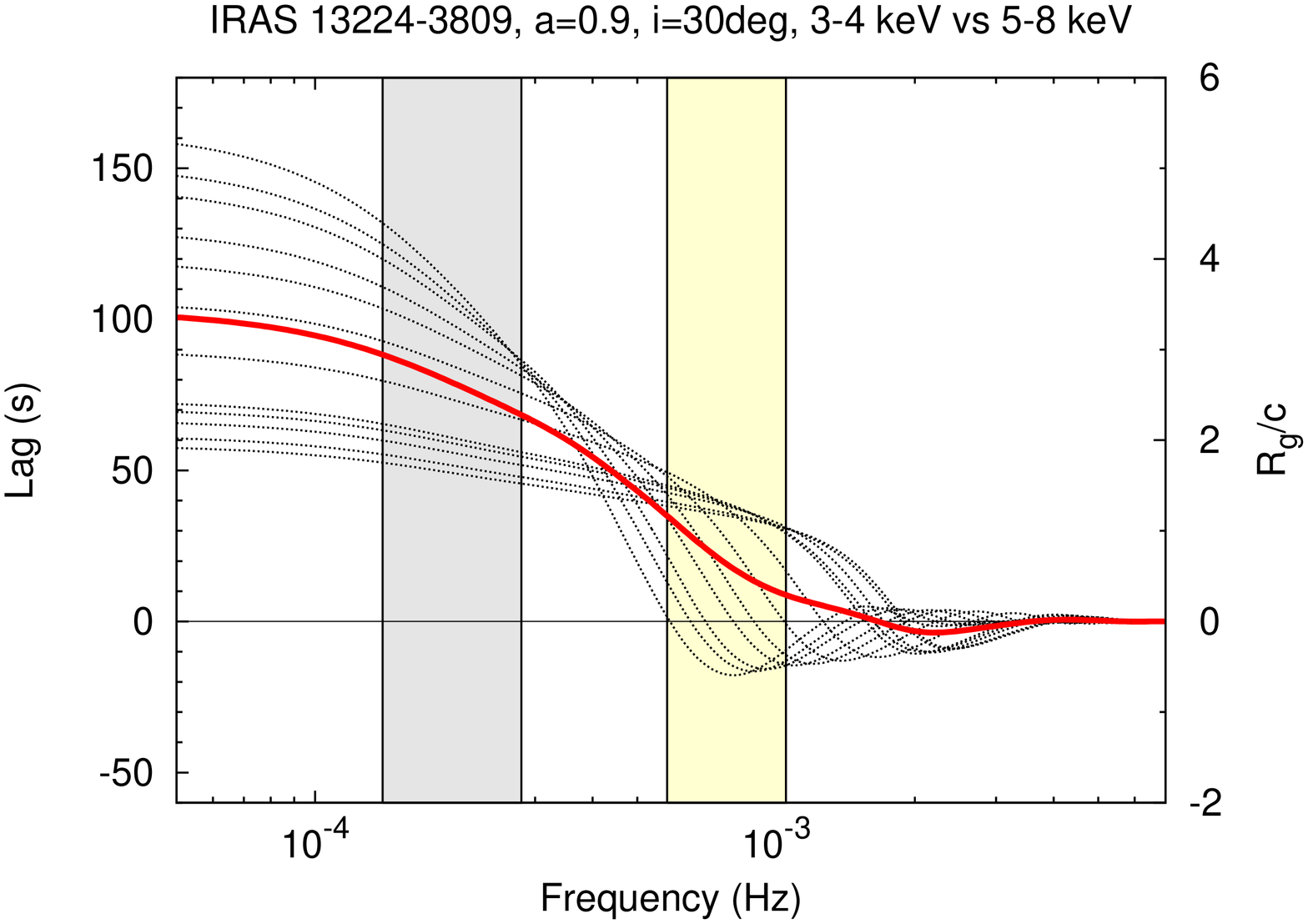}}
\resizebox{8cm}{!}{\includegraphics[width=65mm,angle=270]{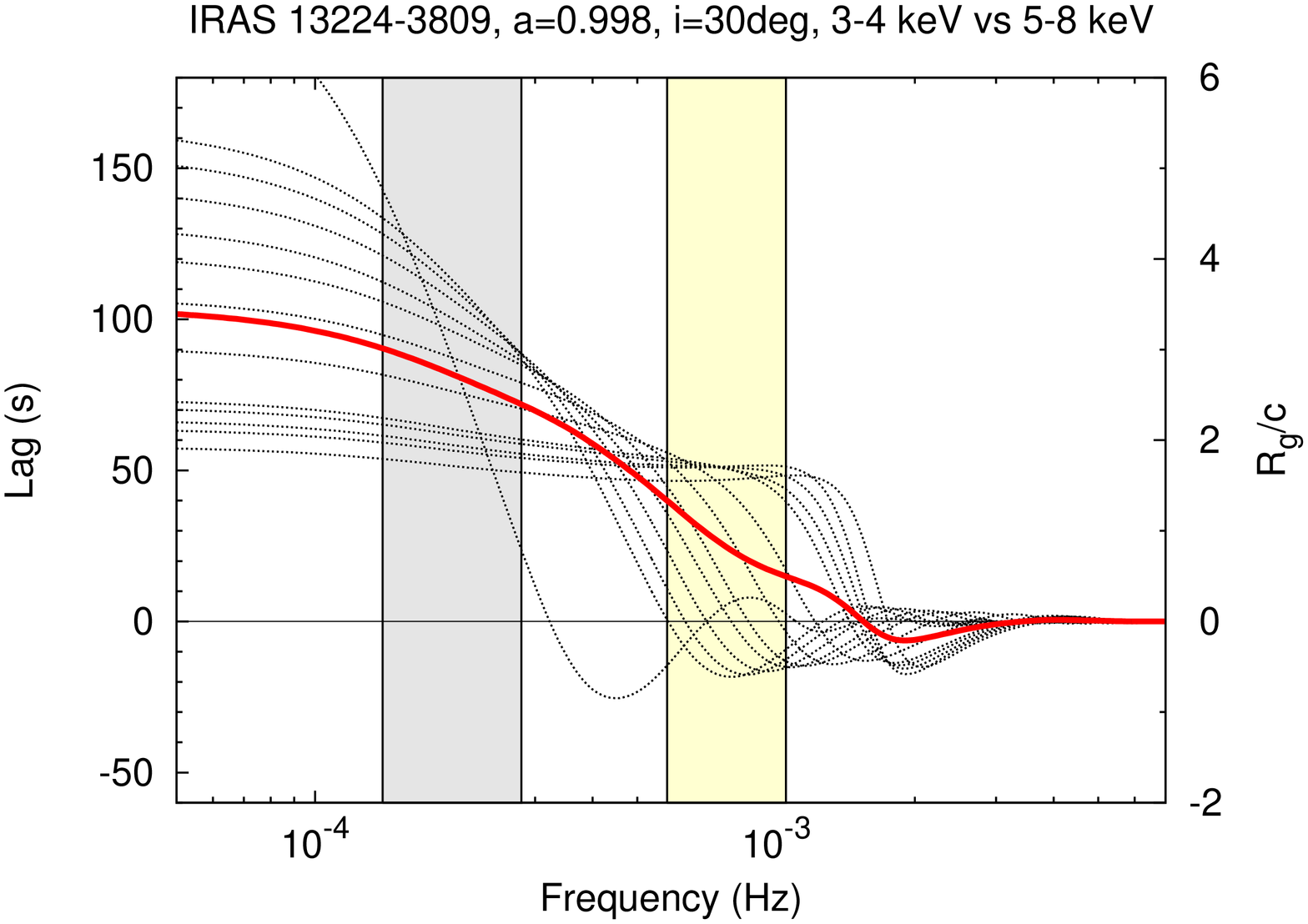}}
}\caption{
Same as figure \ref{fig:lagf}, but for $i=30$~deg
}
\label{fig:lagf_30}
\end{center}
\end{figure*}

Figure \ref{fig:ironspectra2} shows $h$ dependence of the Fe-K line profiles and fluxes for different $a$.
Doppler effect becomes weaker and thus a narrower Fe line is produced than the $i=60$~deg cases;
the blue cut-off energy and the red horn energy are $\sim6.8$~keV and $\sim5.8$~keV, respectively, 
whereas they are $\sim8$~keV and $\sim5.2$~keV for $i=60$~deg.
In particular, $h$ dependence of the line flux is very different;
the peak flux of the iron line varies by one order of magnitude between $h=3$ and 8 for $i=30$~deg,
whereas it is only a factor of 2 for $i=60$~deg.
The large RDC variability makes the EW plots rather flat.
The rms dip is not produced in this case;
on the contrary, change of the line profile makes narrow {\it peaks} around 7~keV (figure \ref{fig:rms_30}).
Such peaks were predicted in model A of \citet{nie10};
they showed that increase of fractional variability in the Fe-K energy band is produced especially in the low-inclination cases ($i\leq40$~deg).
In this case, change of the Fe-K spectral shape depending on $h$ is larger than the flux variability,
and the variability amplitude rather increases.
This rms peak brings us a strong statement that the relativistic light bending model has to be largely modified.

\begin{figure}
\centering
\includegraphics[width=70mm,angle=270]{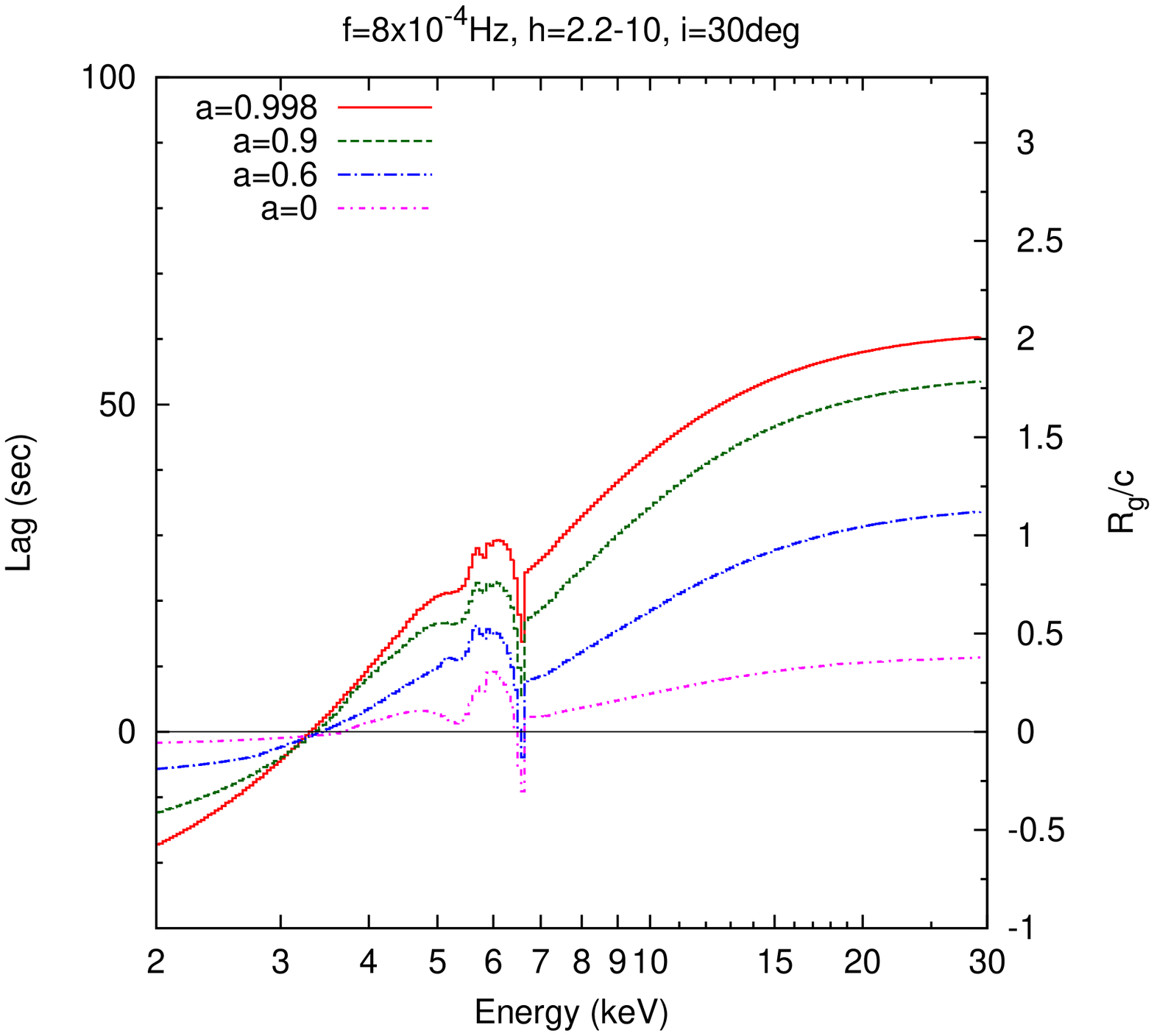}
\caption{
Same as figure \ref{fig:lagE}, but for $i=30$~deg
}
\label{fig:lagE_30}
\end{figure}

The 2D transfer functions, the lag-frequency plot, and the lag-energy spectrum have 
similar tendency to those for $i=60$~deg, with small differences (figures \ref{fig:2dtf_30}, \ref{fig:lagf_30}, \ref{fig:lagE_30}).
The lag-frequency plots look very similar, and
the Fe-K features in the lag-energy plots are sharper and less redshifted,
which reflects the line profiles in the energy spectra.

%%%%%%%%%%%%%%%%%%%%%%%%%%%%%%%%%%%%%%%%%%%%%%%%%%

\end{document}